\documentclass[prd,byrevtex
,tightenlines
,showkeys
,showpacs]{revtex4}

\usepackage{amsmath,amsfonts,amssymb}
\usepackage[pdftex]{graphicx}
\usepackage{float}
\newcommand{\bs}[1]{\boldsymbol{#1}}

\newcommand{\al}{\alpha}

\newcommand{\ep}{\epsilon}

\begin{document}
\title{Entanglement dynamics in  de Sitter spacetime}
\author{Shingo Kukita}
\email{kukita@th.phys.nagoya-u.ac.jp}
\author{Yasusada Nambu}
\email{nambu@gravity.phys.nagoya-u.ac.jp}
\affiliation{Department of Physics, Graduate School of Science, Nagoya 
University, Chikusa, Nagoya 464-8602, Japan}

\date{October 16, 2017 ver 1.3} 
\begin{abstract}
  We apply the master equation with dynamical coarse graining approximation
  to a pair of detectors interacting with a scalar field. By solving
  the master equation numerically, we investigate evolution of
  negativity between comoving detectors in de Sitter space. For the
  massless conformal scalar field with the conformal vacuum, it is
  found that a pair of detectors can perceive entanglement beyond the
  Hubble horizon scale if the initial separation of detectors is
  sufficiently small. At the same time, violation of the Bell-CHSH
  inequality on the super horizon scale is also detected. For the
  massless minimal scalar field with the Bunch-Davies vacuum, on the
  other hand, the entanglement decays within Hubble time scale owing
  to the quantum noise caused by particle creations in de Sitter space
  and the entanglement on the super horizon scale cannot be detected.
\end{abstract}
\maketitle
\section{Introduction}
 
The entanglement is one of the most important quantity in the quantum
mechanics and quantum information~\cite{Nielsen2007}. In the context
of the inflationary scenario, investigating nature of entanglement for
quantum fields in de Sitter spacetime is crucial to understand
mechanism of quantum to classical transition of primordial fluctuation
generated during
inflation~\cite{Nambu2008,Nambu2009,Nambu2011,Nambu2013}.  This
subject is also expected to reveal the relation between causal
structures of spacetime and quantum information.  In cosmological
situation, the particle detector model
\cite{Unruh1976a,DeWitt1979,Birrell1984} is applied to study nature of
entanglement of quantum fields. In de Sitter space, with the lowest
order perturbation calculation, it was shown that detection of
entanglement of the quantum field is possible when a separation of a
pair of detectors is smaller than $2H^{-1}$ for the massless conformal
scalar field with the conformal vacuum and $0.6H^{-1}$ for the minimal
massless scalar field with the Bunch-Davies vacuum
\cite{Steeg2009,Nambu2011,Nambu2013,Martin-Martinez2014}. To
investigate time evolution of entanglement between detectors,
perturbative calculation is limited due to appearance of the secular
behavior and it is only possible to discuss initial detection of
entanglement within short time scales.

To explore long time dynamics of detectors, a method featuring an open
quantum system can be applied~\cite{Breuer2002}: By taking partial
trace of degrees of freedom of the quantum field as a bath
(environment), the dynamics of the detector system is derived as a
quantum master equation.  In general, the master equation becomes an
integro-differential equation, and the state at the specific time
depends on its past evolution (i.e. it is non-Markovian
nature). However, it is possible to recover the Markovian property of
the master equation by assuming that the bath time scale is
sufficiently shorter than the relaxation time scale of the system. By
combining this assumption with so-called the secular approximation
(rotation wave approximation), which neglects transition via
non-energy-conversing processes, it can be shown that the resulting
master equation has the Gorni-Kossakowski-Lindblad-Sudarshan
(GKLS)~\cite{Gorini1976,Lindblad1976} form, and preserves the trace and
the complete positivity of the state in course of time evolution.

Many authors have applied the quantum master equation to explore
physics of quantum field in curved spacetimes.  A single detector
system was investigated to study thermalization by the Hawking
radiation and the Unruh effect~ \cite{Yu2008}, and a two detectors
system was applied to investigate the dynamics of the entanglement for
Unruh effect~\cite{Hu2013,Tian2014,Hu2015,Tian2016}.  In these
previous studies, analyses are relied on the master equations with the
secular approximation.  However, this approximation can not be applied
when the state of the bath is not time translational invariant because
the meaning of energy conservation of the total system is unclear in
such a case. Hence, it is not suitable to apply the master equation
with the secular approximation in cosmological situations.  As an
alternative to the secular approximation, the dynamical coarse
graining (DCG) approximation was proposed
recently~\cite{Lidar2001,Schaller2008,Benatti2010,Majenz2013,Nambu2016a}.
The master equation under this approximation also satisfies the
complete positivity and the Markovian property. It contains a free
parameter $\Delta$ which specifies the coarse graining time scale, and
reduces to the master equation with the secular approximation in the
limit $\Delta \rightarrow \infty$.  By its definition, this master
equation can be applied to the environment with no time translation
invariance such as a quantum field in de Sitter spacetime, provided
that the coarse graining time scale is chosen to be sufficiently
shorter than the Hubble time scale.

To investigate entanglement between two detectors in cosmological
situation, papers \cite{Hu2013,Tian2014} applied the master equation
with the secular approximation. They considered the massless conformal
scalar field in de Sitter spacetime with a static chart and examined
how two static detectors can perceive entanglement of the quantum
field. Due to the Unruh effect and the Hawking radiation in de Sitter
spacetime, they concluded that the entanglement detected has different
behavior from that of the thermal field with Hawking temperature given
by the Hubble parameter in de Sitter space.  In their setting, the
entanglement beyond the Hubble horizon can not be evaluated.  Thus, it
is interesting to investigate how the initial detected entanglement is
affected by quantum noise in de Sitter spacetime after horizon exit.
In this paper, we aim to discuss whether the entanglement between
comoving detectors persist beyond the Hubble horizon.  For this
purpose, we apply a method of the master equation with DCG
approximation to two comoving detectors interacting with a scalar
field in de Sitter space.  We assume two detectors are initially
prepared to be separable and trace evolution of detectors' quantum
state.

The structure of this paper is as follows. In Section II, we introduce
the master equation with DCG approximation and our detector model. In
Section III, we introduce test of the Bell-CHSH inequality in our
set up of detectors model. Then, in Section IV, we investigate the
evolution of the entanglement between two detectors using the master
equation. Section V is devoted to summary.

\section{Master equation of detector model}

\subsection{Master equation with dynamical coarse-graining approximation}

We introduce the quantum master equation with the dynamical coarse
graining approximation
\cite{Lidar2001,Schaller2008,Benatti2010,Majenz2013}. We basically
follow the presentation by Benatti \textit{et
  al.}~\cite{Benatti2010}. The total Hamiltonian is composed of
detector variables (system) interacting with quantum scalar fields
(bath). The total Hamiltonian is
\begin{equation}
H=H_0^{S}+H_0^{B}+ \lambda V,
\end{equation}
where $H_S$ is the system Hamiltonian and $H_B$ is the bath
Hamiltonian and $V=\sum_A\sigma_A\Phi_A$ is the interaction
Hamiltonian, where $\sigma_A$ is the system operator and $\Phi_A$ is
the bath field. $\lambda$ represents a coupling constant and we
assume that interaction between the system and the
bath is small $\lambda \ll 1$ (weak coupling limit).  The total
density operator $\rho_T(t)$ in the Schr\"{o}dinger picture obeys the
von Neumann equation
\begin{equation}
\frac{d\rho_\text{T}(t)}{d t} =
 -i [ H_0+\lambda V, \rho_\text{T}(t) ],\quad H_0=H_0^\text{S}+H_0^\text{B}.
\end{equation}
Our purpose is to obtain an equation for the reduced density
operator for detector system
\begin{equation}
  \rho(t)\equiv\mathrm{Tr}_\text{B}\{\rho_\text{T}(t)\}.
\end{equation}
In the interaction picture, the state is
\begin{equation}
\tilde{\rho}_\text{T}(t)=U(t;t_{0})^\dagger\rho_\text{T}(t)U(t;t_{0}),
\end{equation}
where $U(t;t_{0})$ is the time evolution operator by the free
Hamiltonian $H_0$
\begin{equation}
U(t;t_{0})={\cal T}\exp \{- i \int^{t}_{t_{0}}H_0\,d t' \},
\end{equation}
and ${\cal T}$ denotes the time ordering operator.

Then, $\tilde{\rho}_\text{T}(t)$ obeys
\begin{equation}
\frac{d\tilde{\rho}_\text{T}}{dt}=
 -i [ \lambda \tilde{V}(t), \tilde{\rho}_\text{T} ]
,\quad
\tilde{V}(t)=U(t;t_{0})^\dagger V U(t;t_{0}).
\end{equation}
The solution of the above equation up to the second order of the
perturbation is
\begin{align}
\tilde{\rho}_\text{T}(t)&=
\tilde{\rho}_\text{T}(t_0)
- i\lambda \int_{t_0}^{t}dt_1
         \,[\tilde{V}(t_1),  \tilde{\rho}_\text{T}(t_0)]\nonumber\\
&-\lambda^{2}\int_{t_0}^{t}dt_1\int_{t_0}^{t_1}dt_2
\, [\tilde{V}(t_1), [\tilde{V}(t_2),
 \tilde{\rho}_\text{T}(t_0)]]+O(\lambda^3).
\label{eq:naive}
\end{align}
We assume that during evolution, the state of the total system is a
product state
\begin{equation}
\tilde{\rho}_\text{T}(t)\approx\tilde{\rho}(t)\otimes \rho_{B}.
\end{equation}
where $\rho_B=\rho_B(t_0)$. This assumption is justified because the
interaction between the system and the bath is weak and the
correlation between the system and the bath can be neglected when the
bath time scale is shorter than the system time scale. By taking the
trace of the perturbative solution \eqref{eq:naive} with respect to
the bath degrees of freedom,
\begin{equation}
\tilde{\rho}(t)=
\tilde{\rho}(t_0)-\lambda^{2}\int_{t_0}^{t}dt_1
\int_{t_0}^{t_1}dt_2
{\mathrm{Tr}}_{B}[\tilde{V}(t_1),
 [\tilde{V}(t_2),
 {\rho}(t_0)\otimes {\rho}_{B}(t_0)]], \label{eq:naive2}
\end{equation}
where we have assumed $\langle\Phi_A\rangle$=0. After rewriting
\eqref{eq:naive2}, 
\begin{equation}
  \tilde\rho(t)-\tilde\rho(t_0)=\Delta
  \times\left(-i[H_{12},\rho_0]+\mathcal{L}[\rho_0]\right),\quad\Delta
  =t-t_0,\quad \rho_0=\rho(t_0),
\end{equation}
and we have defined
\begin{align}
  H_{12}&=-i\frac{\lambda^2}{2\Delta}\sum_{A_1A_2}\int_0^{\Delta
          }ds_1\int_{0}^{\Delta
          }ds_2\,\mathrm{sgn}(s_1-s_2)\,G_{A_1\!A_2}(s_1-s_2)\sigma_{A_1}(s_1+t_0)
          \sigma_{A_2}(s_2+t_0),\\
  \mathcal{L}[\rho_0]&=\frac{\lambda^2}{\Delta
                       }\sum_{A_1A_2}\int_0^{\Delta
                       }ds_1\int_0^{\Delta }ds_2\,G_{A_1\!A_2}(s_1-s_2)\left(
      \sigma_{A_2}(s_2+t_0)\rho_0\sigma_{A_1}(s_1+t_0)-\frac{1}{2}
      \{\sigma_{A_1}(s_1+t_0)\sigma_{A_2}(s_2+t_0),\rho_0\}\right),
\end{align}
where the bath correlation function is introduced by
\begin{equation}
  G_{A_1\!A_2}(t_1-t_2)=\mathrm{Tr}_B\{\Phi_{A_1}(t_1)\Phi_{A_2}(t_2)\rho_B(t_0)\}
  =\langle\Phi_{A_1}(t_1)\Phi_{A_2}(t_2)\rangle.
\end{equation}
The time dependence of the system variable can be written as
\begin{equation}
  \sigma_A(t)=\sum_Bu_{AB}(t-t_0)\sigma_B,\quad u_{AB}(0)=1,
\end{equation}
where the specific form of the function $u_{AB}(t)$ depends on the
system Hamiltonian. Using this relation, we obtain
\begin{align}
  H_{12}&=\sum_{B_1B_2}H_{B_1\!B_2}\sigma_{B_1}\sigma_{B_2},\\
  \mathcal{L}[\rho_0]&=\sum_{B_1B_2}C_{B_1\!B_2}\left(
           \sigma_{B_2}\rho_0\,\sigma_{B_1}-\frac{1}{2}\{\sigma_{B_1}\sigma_{B_2},
                       \rho_0\}\right),
\end{align}
with
\begin{align}
  &H_{B_1\!B_2}=-i\frac{\lambda^2}{2\Delta}\sum_{A_1A_2}\int_0^{\Delta
    }ds_1\int_0^{\Delta
    }ds_2\,\mathrm{sgn}(s_1-s_2)\,G_{A_1\!A_2}(s_1-s_2)
    u_{A_1\!B_1}(s_1)u_{A_2\!B_2}(s_2),\\
  &C_{B_1\!B_2}=\frac{\lambda^2}{\Delta}\sum_{A_1A_2}\int_0^{\Delta
    }ds_1\int_0^{\Delta}ds_2\,G_{A_1\!A_2}(s_1-s_2)\,u_{A_1\!B_1}(s_1)u_{A_2\!B_2}(s_2).
\end{align}
As $t_0$ is an arbitrary initial time in Eq.~\eqref{eq:naive2}, it is
possible to convert \eqref{eq:naive2} to the master equation by
assuming that the bath time scale is shorter than the system time
scale, and the interaction is weak \cite{Benatti2010,Nambu2016a}. In
the Schr\"{o}dinger picture,
\begin{equation}
  \frac{d}{dt}\rho=-i[H_\text{eff},\rho]+\mathcal{L}[\rho],\quad
  H_\text{eff}=H_0^S+H_{12}.
  \label{eq:master}
\end{equation}
This is the master equation with dynamical coarse-graining
approximation and has the GKLS form. This time scale of
coarse graining is specified by the parameter $\Delta$. It can be
shown that the coefficients $C_{B_1\!B_2}$ in $\mathcal{L}[\rho]$ form
a positive matrix. Thus, this master equation preserves the trace and
the complete positivity of the state. In the limit of
$\Delta\rightarrow\infty$,  \eqref{eq:master}
reduces to the master equation with the secular approximation.

\subsection{Particle detector model}
We present explicit form of the master equation for a two detectors
interacting with a scalar field. Detectors are assumed to have two
internal energy levels. The Hamiltonian of the total system is
\begin{equation}
H=H_0^S+\lambda
V+H_\phi=\sum_{\al=1,2}\frac{\omega}{2}\sigma_{3}^{(\al)}+\lambda\sum_{\al=1,2}
(\sigma_{+}^{(\al)}+\sigma_{-}^{(\al)})
\,\phi(\bs{x}_\al)+H_\phi
\end{equation}
where $\sigma^{(1)}_{i}=\sigma_{i}\otimes\openone,
\sigma^{(2)}_{i}=\openone\otimes \sigma_{i},
i=3,\pm$ and $\sigma_3=\begin{pmatrix} 1 & 0 \\ 0 & -1 \end{pmatrix},
\sigma_{+}=\begin{pmatrix} 0 & 1 \\ 0 & 0 \end{pmatrix},
\sigma_{-}=\begin{pmatrix} 0 & 0 \\ 1 & 0
\end{pmatrix}$ with the basis of a detector $\{|1\rangle, |0\rangle\}$.
$\bs{x}_{\al}$
denotes position of detectors. We assume two detectors have no direct
interaction. For this Hamiltonian, the coefficients appear in the
master equation \eqref{eq:master} are
\begin{align}
H_{\text{eff}}&= H_0^{S} -\frac{i}{2}\sum_{\al_1, \al_2 = 1}^{2}
\sum_{j_1,j_2=\pm}H_{j_1j_2}^{|\al_1-\al_2|}\sigma^{(\al_1)}_{j_1}\sigma^{(\al_2)}_{j_2
                },
\\
\mathcal{L}[\rho] &= \frac{1}{2}\sum_{\al_1, \al_2 = 1}^{2} \sum_{j_1,
  j_2=\pm}C_{j_1j_2}^{|\al_1-\al_2|}\left[2\sigma^{(\al_2)}_{j_2}\rho\,
\sigma^{(\al_1)}_{j_1}
-\sigma^{(\al_1)}_{j_1}
\sigma^{(\al_2)}_{j_2}\rho-\rho\,\sigma^{(\al_1)}_{j_1}
\sigma^{(\al_2)}_{j_2}\right],
\end{align}
with
\begin{align}
C^{|\al_1-\al_2|}_{j_{1}j_{2}}&=\frac{\lambda^{2}}{\Delta}
                                    \int^{\Delta }_{0}ds_{1}ds_{2}\,
 e^{i\omega(j_1s_1+j_2s_2)}D(\bs{x}_{\al_1},t+s_{1};\bs{x}_{\al_2},t+s_{2}),\\
H^{|\al_1-\al_2|}_{j_1j_2}&=
\frac{\lambda^{2}}{\Delta} \int^{\Delta}_{0}ds_{1}ds_{2}\,
 \mathrm{sgn}(s_{1}-s_{2})
      e^{i\omega(j_1s_1+j_2s_2)}D(\bs{x}_{\al_1},t+s_{1}
;\bs{x}_{\al_2},t+s_{2}),
\end{align}
and
$D({\bs x}_{\al_1},t_{1};{\bs
  x}_{\al_2},t_{2})=\langle\phi(\bs{x}_{\al_1},t_1)\phi(\bs{x}_{\al_2},t_2)\rangle$
is the Wightman function of the scalar field, $\Delta$ is a time
coarse-graining parameter. As the master equation with DCG
approximation is based on the assumption of the stationarity of the
bath, we must impose $\Delta<H^{-1}$ for the cosmological situation.

As a quantum field, we consider the massless conformal  scalar field
and the massless minimal scalar field in de Sitter space with a flat
spatial slice. The Wightman function for the conformal massless scalar
with the conformal vacuum is
\begin{equation}
D_\text{conf}(r_{12},X,Y)=
-\frac{H^{2}}{16\pi^{2}}\frac{1}{\sinh^{2}(HY-i\epsilon)-e^{2HX}(Hr_{12}/2)^{2}},
\end{equation}
where $X=(t_{1}+t_{2})/2$, $Y=(t_{1}-t_{2})/2$ and
$r_{12}=|{\bs x}_{\al_1}-{\bs x}_{\al_2}|$. For the massless minimal scalar field
with Bunch-Davies vacuum state,
\begin{align}
  &D_\text{min}=D_\text{conf}+D_2, \notag\\
  &D_2=-\frac{H^2}{8\pi^2}\Bigl\{\mathrm{Ei}\left[-\frac{ik_0}{H}\left(
    -Hr_{12}+2e^{-HX}\sinh(H(Y-i\ep))\right)\right]+
\mathrm{Ei}\left[\frac{ik_0}{H}\left(
    -Hr_{12}+2e^{-HX}\sinh(H(Y-i\ep))\right)\right]\Bigr\}+\frac{H^2}{4\pi^2},
\end{align}
where $k_0$ is an infrared cutoff corresponding to the initial size of
the inflating universe and we take this value as $k_0=H$ in our
analysis.  $\mathrm{Ei}(-x)=-\int_x^\infty \frac{dy}{y}e^{-y}$ is the
exponential integral. For a technical reason to evaluate integrals
contained in coefficients of the master equation, we replace the
integrals with the following Gaussian form
\begin{equation}
\int^{\Delta}_{0}ds_1ds_2 \rightarrow
\frac{2}{\pi}\int^{\infty}_{-\infty}ds_1ds_2
\exp
\left[-\frac{(s_1-\sigma)^2}{2\sigma^2}-\frac{(s_2-\sigma)^2}{2\sigma^2}
\right],\quad \sigma=\Delta/2.
\label{int}
\end{equation}
Then, by introducing new integration variables as $x=(s_1+s_2)/2,
y=(s_1-s_2)/2$, the coefficients becomes 
\begin{align}
  &C_{j_1j_2}^{|\al_1-\al_2|}=\frac{2\lambda^2}{\pi\sigma}e^{-\omega^2\sigma^2}
    e^{i\omega\sigma j_{+}}\int_{-\infty}^{+\infty} dxdy\,
    e^{-\frac{1}{\sigma^2}[x-(\sigma+\frac{i}{2}\omega\sigma^2j_{+})]^2}e^{-\frac{1}{\sigma^2}(y-\frac{i}{2}\omega\sigma^2j_{-})^2}D(r_{12},t+x,y),\\
  &H_{j_1j_2}^{|\al_1-\al_2|}=\frac{2\lambda^2}{\pi\sigma}e^{-\omega^2\sigma^2}
    e^{i\omega\sigma j_{+}}\int_{-\infty}^{+\infty} dxdy\,\mathrm{sgn}(y)\,
    e^{-\frac{1}{\sigma^2}[x-(\sigma+\frac{i}{2}\omega\sigma^2j_{+})]^2}e^{-\frac{1}{\sigma^2}(y-\frac{i}{2}\omega\sigma^2j_{-})^2}D(r_{12},t+x,y),
\end{align}
with $j_{\pm}=j_1\pm j_2$.  To evaluate these integrals, we apply the
saddle point approximation after shifting the contour of integrals in
complex $x$ and $y$ planes. After evaluating $x$ integrals by the
saddle point approximation, 
\begin{align}
  &C_{j_1j_2}^{|\al_1-\al_2|}=\frac{2\lambda^2}{\pi^{1/2}}e^{-\omega^2\sigma^2}
    e^{i\omega\sigma j_{+}}\int_{-\infty}^{+\infty} dy\,
    e^{-\frac{1}{\sigma^2}(y-\frac{i}{2}\omega\sigma^2j_{-})^2}
D(r_{12},t+\sigma+\frac{i}{2}\omega\sigma^2j_{+},y),\\
  &H_{j_1j_2}^{|\al_1-\al_2|}=\frac{2\lambda^2}{\pi^{1/2}}e^{-\omega^2\sigma^2}
    e^{i\omega\sigma j_{+}}\int_{-\infty}^{+\infty}dy\,\mathrm{sgn}(y)\,
    e^{-\frac{1}{\sigma^2}(y-\frac{i}{2}\omega\sigma^2j_{-})^2}
    D(r_{12},t+\sigma+\frac{i}{2}\omega\sigma^2j_{+},y).
\end{align}
For $y$ integrals, we must require the following conditions for
parameters contained in the master equation:
\begin{itemize}
  \item The shift of contours does not cross poles of the Wightman
  function: $H\omega\sigma^2<\pi$.
  \item The width of the Gaussian factor $\sigma$ is smaller than the
  shift of contours: $\sigma<\omega\sigma^2$.
  \item The width of the Gaussian factor $\sigma$ is smaller than
  separation of poles of the Wightman function: $\sigma<\pi/H,
  \sigma<r$.
  \item The stationarity of the bath: $\sigma<H^{-1}$.
\end{itemize}
Combining these conditions, we have the following constraints for
parameters of the master equation under the saddle point approximation
\begin{equation}
  \sigma H<1,\quad 1<\omega\sigma,\quad \omega\sigma^2<\pi/H,\quad \sigma<r.
\end{equation}
We present explicit form of coefficients of the master equation for the
minimal scalar field case.
\subsubsection{Coefficients $C_{j_1j_2}^{|\al_1-\al_2|}$}
\begin{figure}[H]
  \centering
  \includegraphics[width=0.4\linewidth,clip]{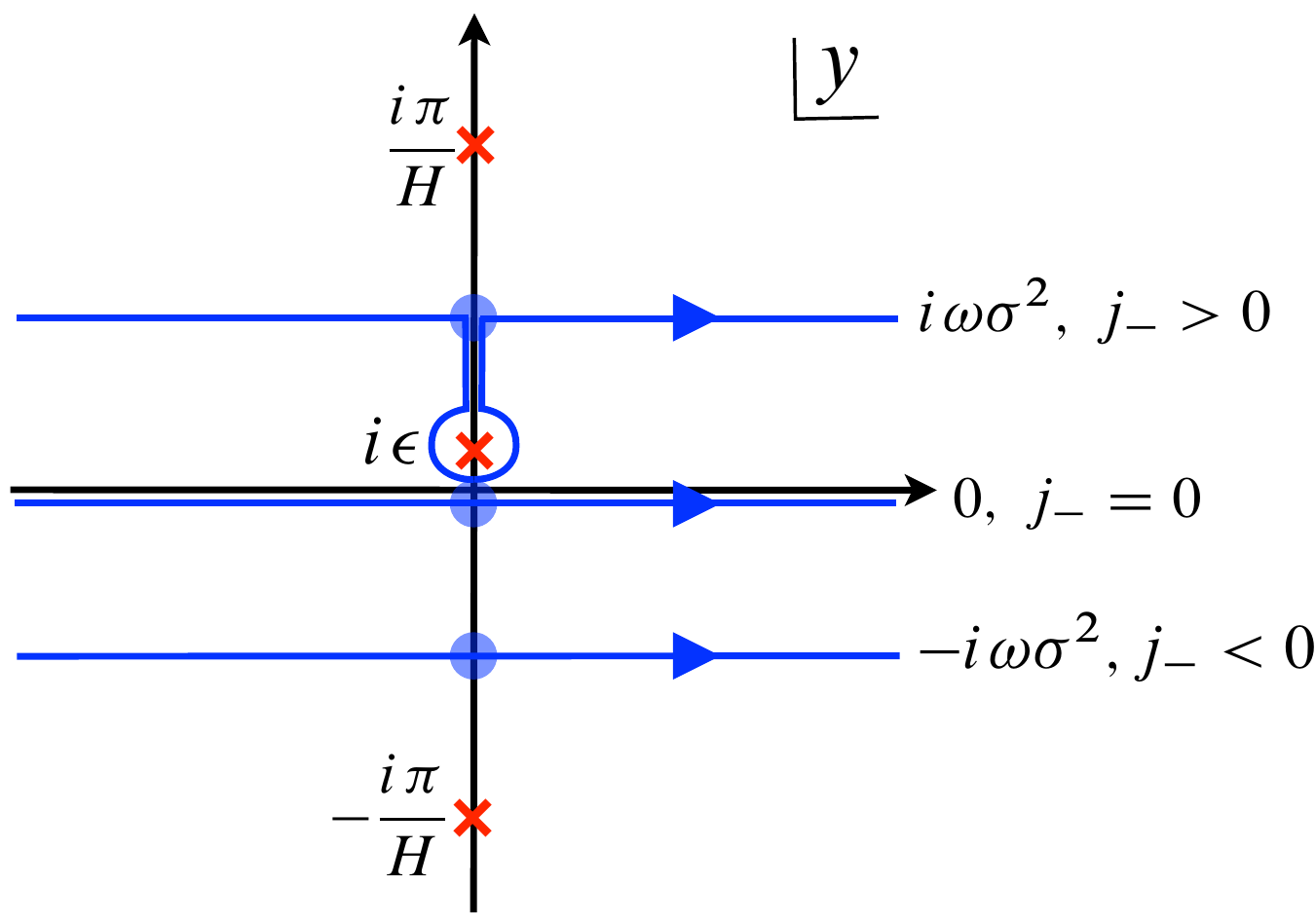}%
  \hspace{0.7cm}
  \includegraphics[width=0.4\linewidth,clip]{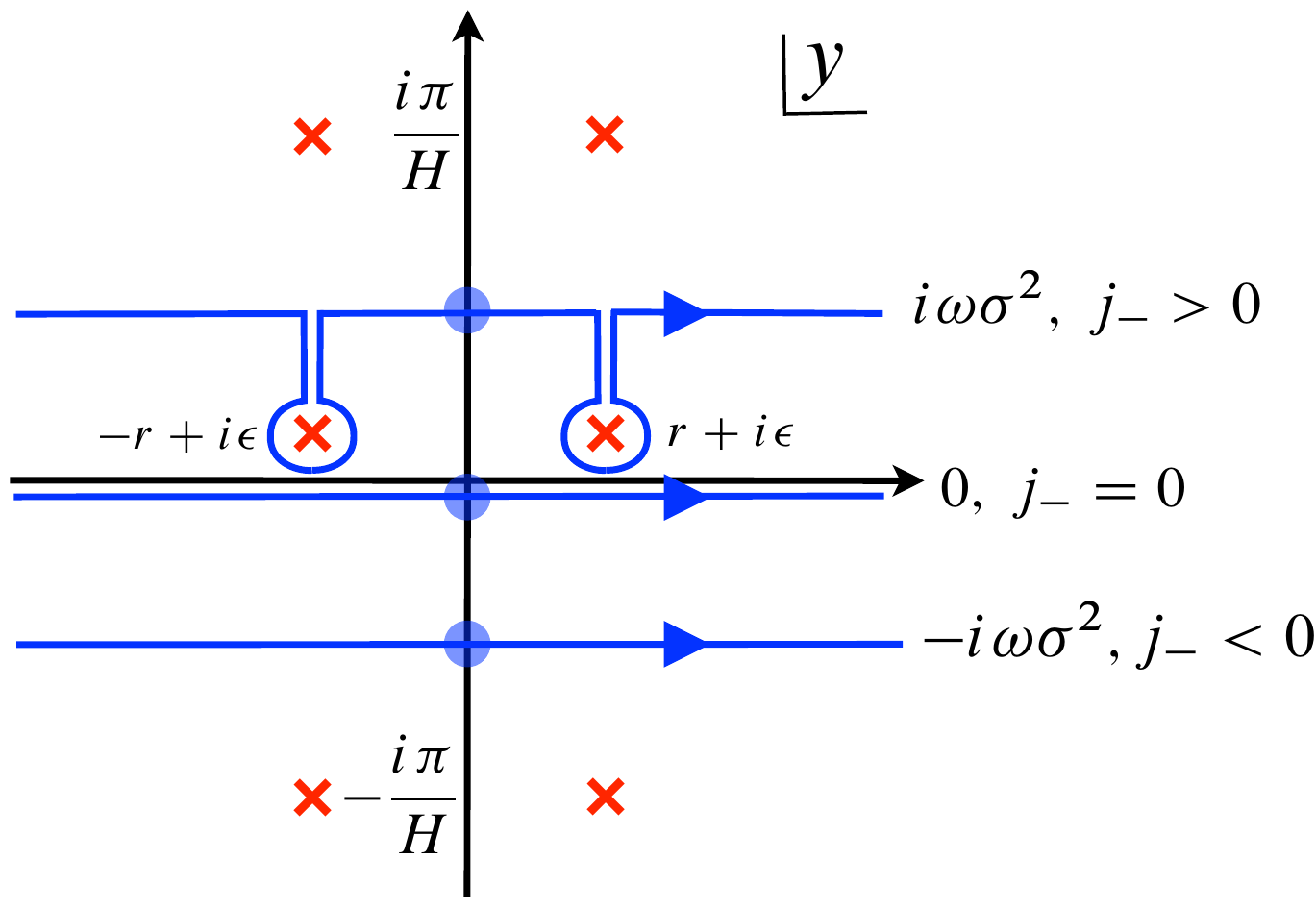}
  \caption{Contours of $y$ integration for
    $C_{j_1j_2}^{|\al_1-\al_2|}$. Cross symbols represent poles of
    $D_{\text{conf}}$ and disks represent saddle points of
    integrand. Left panel: $\al_1=\al_2~(r_{12}=0)$, Right panel
    : $\al_1\neq\al_2~(r_{12}=r\neq 0)$.}
  \label{fig:cont1}
\end{figure}

We must consider contribution of poles in $D_\text{conf}$. For
$\al_1=\al_2~ (r_{12}=0)$, the contours of integration are shown in
Fig.~\ref{fig:cont1} (left panel).
\noindent
The result for $\al_1=\al_2~(r_{12}=0)$ is
\begin{equation}
  C_{j_1j_2}^0=2\lambda^2e^{-\omega^2\sigma^2}\times
  \begin{cases}
    e^{i\omega\sigma j_{+}}\left(\sigma
      D_2(r_{12}=0, x=t+\sigma+\frac{i}{2}\omega\sigma^2 j_{+},y=0)+\frac{1}{8\pi^2\sigma}\right)&\quad
      (C_{++}^0, C_{--}^0) \\
      \sigma D_\text{min}(r_{12}=0,x=t+\sigma,y=i\omega\sigma^2)+\frac{\omega
        e^{\omega^2\sigma^2}}{4\pi^{1/2}}&\quad (C_{+-}^0)\\
      \sigma D_\text{min}(r_{12}=0,x=t+\sigma,y=-i\omega\sigma^2)&\quad (C_{-+}^0)
  \end{cases}
\end{equation}
where $1/(8\pi^2\sigma)$ and $\omega
e^{\omega^2\sigma^2}/(4\pi^{1/2})$ are contributions from the pole
$i\ep$ in $D_\text{conf}$.

For $\al_1\neq\al_2~(r_{12}=r\neq 0)$,
\begin{equation}
  C_{j_1j_2}^1=2\lambda^2e^{-\omega^2\sigma^2}\times
  \begin{cases}
    e^{i\omega\sigma j_{+}}\sigma
    D_\text{min}(r,t+\sigma+\frac{i}{2}\omega\sigma^2j_{+},y=0)&\quad (C_{++}^1,
    C_{--}^1)\\
    \sigma D_\text{min}(r,t+\sigma,i\omega\sigma^2)+2i\pi^{1/2}\mathrm{Res}\left[
      e^{-(y-i\omega\sigma^2)^2/\sigma^2}D_\text{conf}(r,t+\sigma,y)\right]&\quad
    (C_{+-}^1)\\
    \sigma D_\text{min}(r,t+\sigma,-i\omega\sigma^2)&\quad (C_{-+}^1)
  \end{cases}
\end{equation}
where the residue (sum of contributions of two poles) is
\begin{equation}
  \mathrm{Res}[\cdots]=\frac{1-\exp\left[\frac{4i\omega}{H}\sinh^{-1}
\left(e^{H\sigma}Hr_p/2\right)\right]}{8\pi^2r_p(4+e^{2H\sigma}(Hr_p)^2)^{1/2}}
\times\exp\left[-H\sigma+\left(\omega\sigma-\frac{i}{H\sigma}\sinh^{-1}(e^{H\sigma}Hr_p/2)\right)^2\right],\quad r_p=re^{Ht}.
\end{equation}

\subsubsection{Coefficients $H_{j_1j_2}^{|\al_1-\al_2|}$}
\begin{figure}[H]
  \centering
  \includegraphics[width=0.4\linewidth,clip]{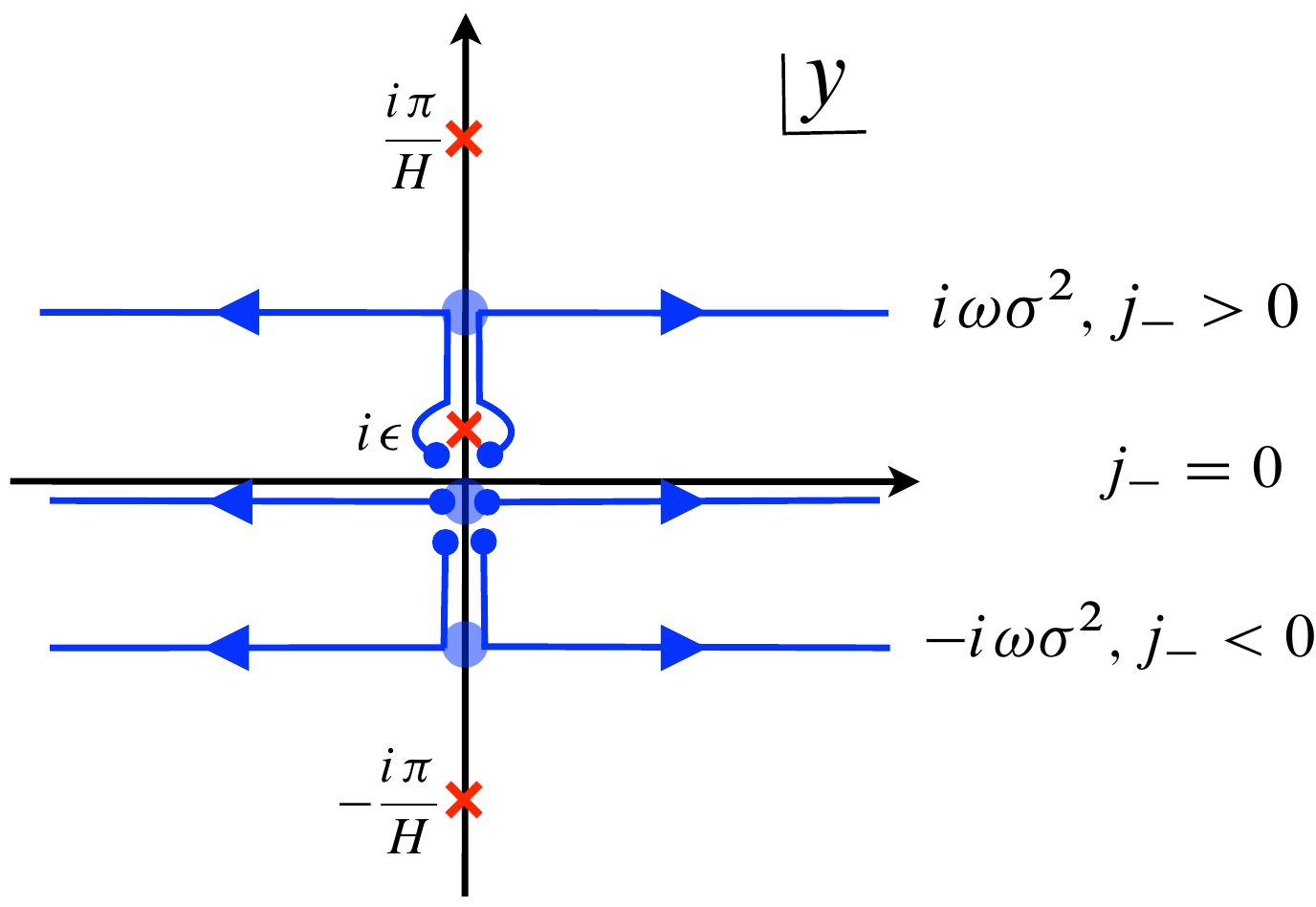}%
  \hspace{0.7cm}
  \includegraphics[width=0.4\linewidth,clip]{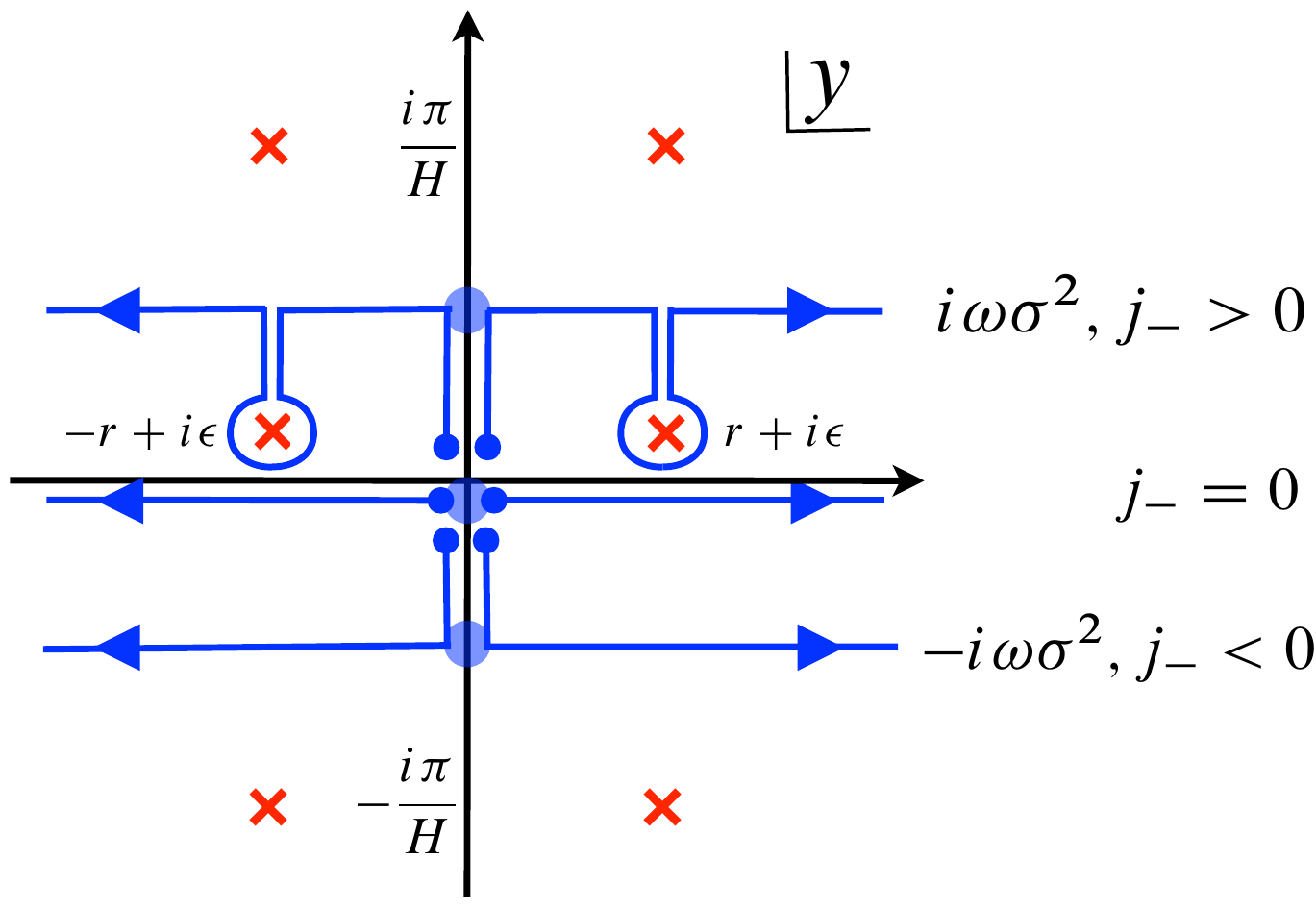}
  \caption{Contours of $y$ integration for
    $H_{j_1j_2}^{|\al_1-\al_2|}$. Left panel: $\al_1=\al_2~(r_{12}=0)$,
    Right panel: $\al_1\neq\al_2~(r_{12}=r\neq 0)$.}
  \label{fig:cont2}
\end{figure}

The contours of integration is shown in Fig.~\ref{fig:cont2}.
\noindent
For $\al_1=\al_2$, contribution from saddles to the integrals cancels out
and we only have contribution from a pole and contours along the
imaginary $y$ axis:
\begin{equation}
  H_{j_1j_2}^0=
  \begin{cases}
    \frac{1}{\ep}+I_1&\quad (H_{+-}^0)\\
    0&\quad (H_{++}^0, H_{--}^0)\\
    -I_1&\quad (H_{-+}^0)
  \end{cases}
\end{equation}
where $I_1$ is a finite contribution from contours along
$\mathrm{Im}(y)$ axis and its explicit form is not necessary in the
present analysis.  The coefficients $H_{+-}^0$ diverges as
$\ep\rightarrow 0$ and we will see this divergence can be renormalized
to $\omega$ (Lamb shift). For $\al_1\neq\al_2~(r_{12}=r\neq 0)$,
\begin{equation}
  H_{j_1j_2}^1=
  \begin{cases}
    4i\pi^{1/2}\lambda^2e^{-\omega^2\sigma^2}e^{i\omega\sigma j_{+}}
    \left(\mathrm{Res}_1-\mathrm{Res}_2\right)+I_2 &\quad (H_{+-}^1)\\
    0 &\quad (H_{++}^1, H_{--}^1)\\
    -I_2 &\quad (H_{-+}^1)
  \end{cases}
\end{equation}
where $\mathrm{Res}_{1,2}$ denotes residues of two poles and $I_2$
is a contributions from contours along $\mathrm{Im}(y)$ axis, we do not
need their explicit form here.

\subsection{Components of master equation}
We consider the density matrix with the following components
\begin{equation}
  \rho=
  \begin{pmatrix}
    \rho_{11} & 0 & 0 & \rho_{14} \\
    0 & \rho_{22} & \rho_{23} & 0 \\
    0 & \rho_{23}^* & \rho_{33} & 0 \\
    \rho_{14}^* & 0 & 0 & \rho_{44}
  \end{pmatrix},\quad
  \rho_{11}+\rho_{22}+\rho_{33}+\rho_{44}=1, \label{eq:Xstate}
\end{equation}
and we adopt the basis of the state
$\{|11\rangle, |10\rangle, |01\rangle, |00\rangle \}$.  We can check
that this form of the density matrix is compatible with structure of
the master equation and evolution keeps the structure of the matrix 
\eqref{eq:Xstate}. By introducing new symbols for coefficients,
\begin{align*}
  &c^0_{11}=C_{++}^{0}=(C_{--}^{0})^*,\quad
  c^0_{12}=C_{+-}^{0},\quad
  c^0_{21}=C_{-+}^{0},\\
  &c_{11}^1=C_{++}^{1}=(C_{--}^{1})^*,\quad
  c_{12}^1=C_{+-}^{1} ,\quad
  c_{21}^1=C_{-+}^{1}.
\end{align*}
The components of $\mathcal{L}[\rho]$ in the master equation
\eqref{eq:master}  are
\begin{align*}
  &L_{11}=-2 {c^0_{12}}\rho_{11}+{c^0_{21}} \rho_{22}+{c^0_{21}}
    \rho_{33}-{c_{11}^1{}^*}
  \rho_{14}-{c_{11}^1} \rho_{14}{}^*+{c_{21}^1} \rho_{23}+c_{21}^1
  \rho_{23}{}^*,\\
&L_{14}={c^0_{11}}\rho_{23}{}^*+{c^0_{11}}
  \rho_{23}-({c^0_{12}}+{c^0_{21}})\rho_{14}+{c_{11}^1} 
   \left(-\rho_{11}-\rho_{44}+\rho_{22}+\rho_{33}\right)
   ,\\
 &L_{22}={c^0_{12}} \rho_{11}-({c^0_{12}}+{c^0_{21}}) \rho_{22}+{c^0_{21}}
 \rho_{44}-\frac{1}{2}\left(c_{12}^1+c_{21}^1\right)\rho_{23}
-\frac{1}{2}\left(c_{12}^1+c_{21}^1\right) 
   {\rho_{23}{}^*}+{c_{11}^1{}^*}\rho_{14}+{c_{11}^1}
 \rho_{14}{}^*,\\
&L_{23}=c_{12}^1\rho_{11}-\frac{1}{2}(c_{12}^1+c_{21}^1)(\rho_{22}+\rho_{33})
+c_{21}^1\rho_{44}
+c^0_{11}{}^* \rho_{14}+{c_{11}^1}\rho_{14}{}^*-({c^0_{12}}+{c^0_{21}})\rho_{23},\\
&L_{33}={c^0_{12}} \rho_{11}-({c^0_{12}}+{c^0_{21}})
\rho_{33}+{c^0_{21}}\rho_{44}-\frac{1}{2}
\left(c_{12}^1+c_{21}^1\right) \rho_{23}-\frac{1}{2}
\left(c_{12}^1+c_{21}^1\right)\rho_{23}{}^*+{c_{11}^1{}^*}
 \rho_{14}+{c_{11}^1}\rho_{14}{}^*,\\
&L_{41}=-({c^0_{12}}+{c^0_{21}}) \rho_{14}{}^*+c^0_{11}{}^*
  \rho_{23}{}^*+c^0_{11}{}^*
   \rho_{23}+{c_{11}^1{}^*}\left(-\rho_{11}-\rho_{44}+\rho_{22}+\rho_{33}\right)
   ,\\
   &L_{44}={c^0_{12}}\rho_{22}+{c^0_{12}} \rho_{33}-2
     {c^0_{21}}\rho_{44}
-c_{11}^1{}^*
   \rho_{14}-{c_{11}^1}\rho_{14}{}^*+{c_{12}^1} \rho_{23}+{c_{12}^1} \rho_{23}{}^*.
\end{align*}
The components of $i[H_\text{eff},\rho]$ in the master equation
\eqref{eq:master} are
\begin{equation*}
{\small
\begin{pmatrix}
 0  & 0 & 0 & 2i\omega_R\rho_{14} \\
 0 &  (H_{+-}^1+H_{-+}^1) (\rho_{23}^*-\rho_{23})/2 &
    (H_{+-}^1+H_{-+}^1) (\rho_{33}-\rho_{22})/2 & 0 \\
 0 & (H_{+-}^1+H_{-+}^1) (\rho_{22}-\rho_{33})/2 & 
   (H_{+-}^1+H_{-+}^1)(\rho_{23}-\rho_{23}^*)/2 & 0 \\
 - 2i\omega_R\rho_{14}^* & 0 & 0 & 0  \\
\end{pmatrix}
}
\end{equation*}
where $\omega_R=\omega+(H^{0}_{+-}-H^0_{-+})/(2i)$ is a renromalized
frequency. From now on, we denote $\omega_R$ as $\omega$.
In our analysis, we consider the initial state
$\rho_0=|00\rangle\langle00|$. For this initial condition,
$\rho_{22}=\rho_{33}$ and $\mathrm{Im}(\rho_{23})=0$ are kept during time
evolution and we do not have to take into account of $H_{+-}^1+H_{-+}^1$ in
the master equation.

As a measure of entanglement between two detectors, we adopt the
negativity in our analysis. This quantity is introduced via partial
transposition of the  state of \eqref{eq:Xstate}:
$$
 \rho^{PT}=\begin{pmatrix}
   \rho_{11} & 0 & 0 & \rho_{23} \\
   0 & \rho_{22} & \rho_{14} & 0 \\
   0 & \rho_{14}^* & \rho_{33} & 0 \\
   \rho_{23}^* & 0 & 0 & \rho_{44}
   \end{pmatrix}.
$$
The eigenvalues of this state are
$$
 \lambda=\frac{1}{2}\left[(\rho_{22}+\rho_{33})\pm\sqrt{(\rho_{22}-\rho_{33})^2
   +4\rho_{14}\rho_{14}^*}\right],\quad
\frac{1}{2}\left[(\rho_{11}+\rho_{44})\pm\sqrt{(\rho_{11}-\rho_{44})^2
   +4\rho_{23}\rho_{23}^*}\right].
$$
The definition of the negativity is
$$
 E_N\equiv\sum_{\lambda_i<0}|\lambda_i|,
$$
and $E_N>0$ provides a necessary and sufficient condition for
entanglement between two qubits~\cite{Peres1996,Horodecki1996}.  For the initial separable state
$\rho_0=|00\rangle\langle00|$, the solution of the master equation
\eqref{eq:master} around the initial time $t=0$ is
$$
 \rho(t)=\begin{pmatrix}
   0 & 0 & 0 & -c_{11}^1 t \\
   0 & c^0_{21}t & c_{21}^1t& 0 \\
   0 & c_{21}^1t & c^0_{21}t & 0 \\
   -(c_{11}^1)^* t & 0 & 0 & 1-2c^0_{21}t
   \end{pmatrix}.
$$
For this state, the negativity is given by
\begin{equation}
\label{eq:neg0}
 E_N=\max[0, t(|c_{11}^1|-c^0_{21})],\quad t\ll 1.
\end{equation}
This formula is the same as one obtained by perturbation
calculation~\cite{Steeg2009,Nambu2011,Nambu2013,Martin-Martinez2014}
and the value of negativity grows proportional to time.  Initial
separable state instantly becomes entangled if $|c_{11}^1|-c_{21}^0>0$
and detectors can catch entanglement of the quantum field.  Using the
explicit formula of the coefficients,
\begin{equation}
  |c_{11}^1|-c^0_{21}=2\lambda^2\sigma^2e^{-\omega^2\sigma^2}\Bigl(
  \left|D[r,t+\sigma+i\omega\sigma^2,0]\right|
  -D[0,t+\sigma,-i\omega\sigma^2]\Bigr).
\end{equation}

\section{Bell-CHSH inequality}
We can test violation of the Bell-CHSH inequality in de Sitter space.
For this purpose, we extend our previous analysis of the Bell-CHSH
inequality for detectors system~\cite{Nambu2011}. Let us consider the
following Bell operator:
\begin{equation}
  \mathcal{B}_\text{CHSH}=\bs{a}\cdot\bs{\sigma}\otimes(\bs{b}+\bs{b}')\cdot\bs{\sigma} 
+\bs{a}'\cdot\bs{\sigma}\otimes(\bs{b}-\bs{b}')\cdot\bs{\sigma},
\end{equation}
where $\bs{a},\bs{a}',\bs{b},\bs{b}'$ are real unit vectors. The
Bell-CHSH inequality is
\begin{equation}
  |\langle\mathcal{B}_\text{CHSH}\rangle|\le 2.
\end{equation}
If the state admits a local hidden variable (LHV) description of
correlations, then this inequality holds. Violation of the inequality
means existence of non-locality.  We consider Bloch representation of
the state \eqref{eq:Xstate}:
\begin{align}
  &\rho=\frac{1}{4}\left[I\otimes I+\bs{a}\cdot\bs{\sigma}\otimes
    I+I\otimes
\bs{b}\cdot\bs{\sigma}+\sum_{j,k=1}^3c_{jk}\,\sigma_j\otimes\sigma_k\right],\\
  &\bs{a}=(0,0,\rho_{11}+\rho_{22}-\rho_{33}-\rho_{44}),\quad
    \bs{b}=(0,0,\rho_{11}+\rho_{33}-\rho_{22}-\rho_{44}),\\
  &c_{jk}=\begin{pmatrix}
      2\rho_{23}-2(\rho_{14})_R & 2(\rho_{14})_I & 0 \\
      2(\rho_{14})_I & 2\rho_{23}+2(\rho_{14})_R & 0 \\
      0 & 0 & \rho_{11}+\rho_{44}-\rho_{22}-\rho_{33}
      \end{pmatrix}.
\end{align}
It is known this state violates the Bell-CHSH inequality if and only
if the following condition is satisfied \cite{Horodecki1995a}:
\begin{equation}
    \text{sum of the two largest eigenvalues of the matrix $c\,c^\dagger$~$>1$}
\end{equation}
The eigenvalues of $c\,c^\dagger$ are
\begin{equation}
    4(\rho_{11}+\rho_{44}-1/2)^2,\quad 4(|\rho_{14}|\pm\rho_{23})^2.
\end{equation}
For the initial separable state $\rho_0=|00\rangle\langle00|$, we
cannot expect these eigenvalues exceed unity after evolution because
only non-zero component of the initial state is
$\rho_{44}=1$. However, the Bell-CHSH inequality provides only a
necessary condition for the LHV description and does not guarantee
existence of a LHV~\cite{Gisin1996a}. Hence by passing each detector
through the local filter
\begin{equation}
    f_{A,B}=\begin{pmatrix}1 & 0 \\ 0 & \eta \end{pmatrix},
\end{equation}
there is a possibility revealing hidden non-locality of the
state. After passing through the filter, the state is transformed as
$\rho'=(f_A\otimes f_B)\rho(f_A\otimes f_B)$ and the normalized state
is
\begin{equation}
    \rho'=\frac{1}{\rho_{11}+\eta^2(\rho_{22}+\rho_{33})+\eta^4\rho_{44}}
\begin{pmatrix}
    \rho_{11} & 0 & 0 & \eta^2\rho_{14} \\
    0 & \eta^2\rho_{22} & \eta^2\rho_{23} & 0 \\
    0 & \eta^2\rho_{23} & \eta^2\rho_{33} & 0 \\
    \eta^2\rho_{14}^* & 0 & 0 & \eta^4\rho_{44}
\end{pmatrix}.
\end{equation}
The matrix $c$ becomes
\begin{equation}
    c'=\frac{2\eta^2}{\rho_{11}+\eta^2(\rho_{22}+\rho_{33})+\eta^4\rho_{44}}
\begin{pmatrix}
 \rho_{23}-(\rho_{14})_R & (\rho_{14})_R & 0 \\
 (\rho_{14})_R & \rho_{23}+(\rho_{14})_R & 0 \\
 0 & 0 &
 \frac{\rho_{11}-\eta^2(\rho_{22}+\rho_{33})+\eta^4\rho_{44}}{2\eta^2}
\end{pmatrix}.
\end{equation}
The eigenvalues of $c$ are
\begin{equation}
    \frac{\rho_{11}-\eta^2(\rho_{22}+\rho_{33})+\eta^4\rho_{44}}{
      \rho_{11}+\eta^2(\rho_{22}+\rho_{33})+\eta^4\rho_{44}},\quad
    \frac{2\eta^2(\rho_{23}\pm|\rho_{14}|)}{\rho_{11}+\eta^2(\rho_{22}+\rho_{33})
      +\eta^4\rho_{44}},
\end{equation}
and the condition for violation of the Bell-CHSH inequality is
\begin{align} 
  0>\eta^4-\frac{(\rho_{23}+|\rho_{14}|)^2}{
\rho_{44}(\rho_{22}+\rho_{33})}\eta^2+\frac{\rho_{11}}{\rho_{44}}.
\end{align}
The real parameter $\eta$ satisfying this inequality exists if the
following condition holds
\begin{equation}
    (\rho_{23}+|\rho_{14}|)^4>4\rho_{11}\rho_{44}(\rho_{22}+\rho_{33})^2.
\end{equation}
This inequality provides a sufficient condition for violation of the
Bell-CHSH inequality and if this inequality is satisfied, violation of
the Bell-CHSH inequality can be detected. 

\section{Evolution of entanglement}
We solve the master equation \eqref{eq:master} numerically and follow
evolution of entanglement between two detectors and test violation of
the Bell-CHSH inequality. We used Mathematica and \textsf{NDSolve} to
obtain numerical solutions.

\subsection{Minkowski vacuum}
\begin{figure}[H]
    \centering 
\includegraphics[width=0.4\linewidth,clip]{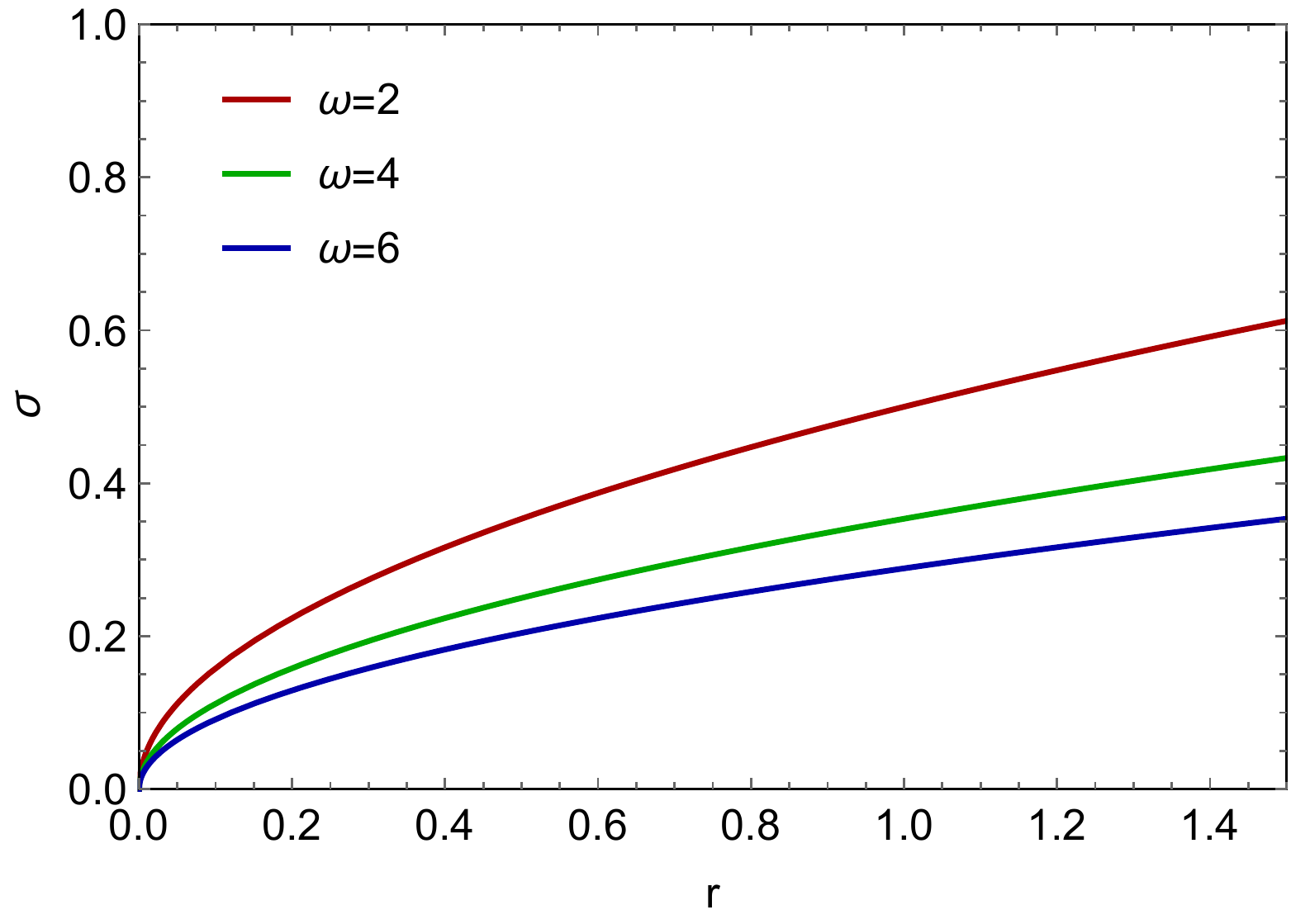}
\caption{The negativity zero lines determined by \eqref{eq:neg0}. For
  parameters in regions above these lines, detectors can detect
  entanglement of the quantum field around initial time.}
    \label{fig:mink-ent0}
\end{figure}

\begin{figure}[H]
    \centering
    \includegraphics[width=0.3\linewidth,clip]{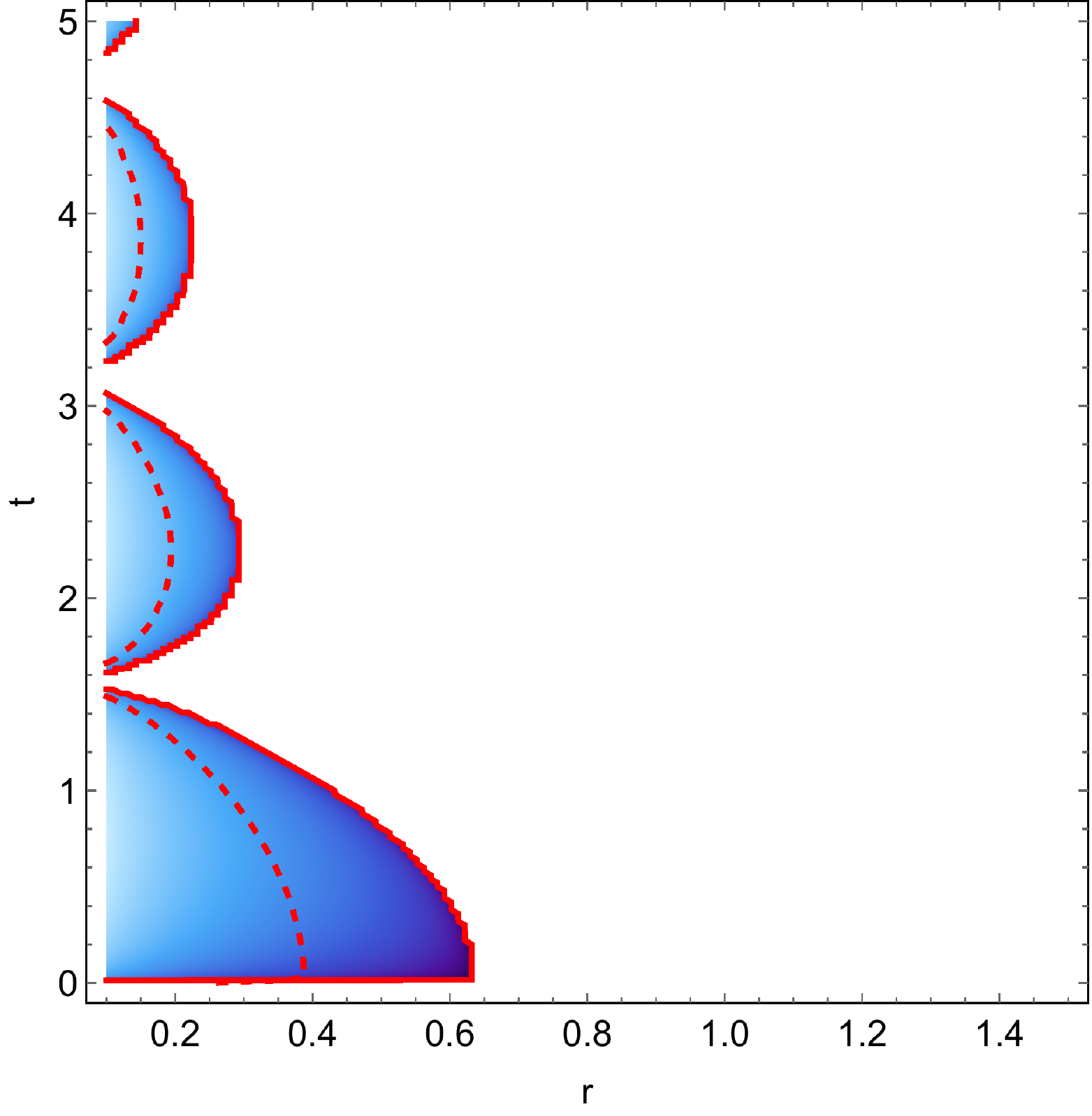}
    \includegraphics[width=0.3\linewidth,clip]{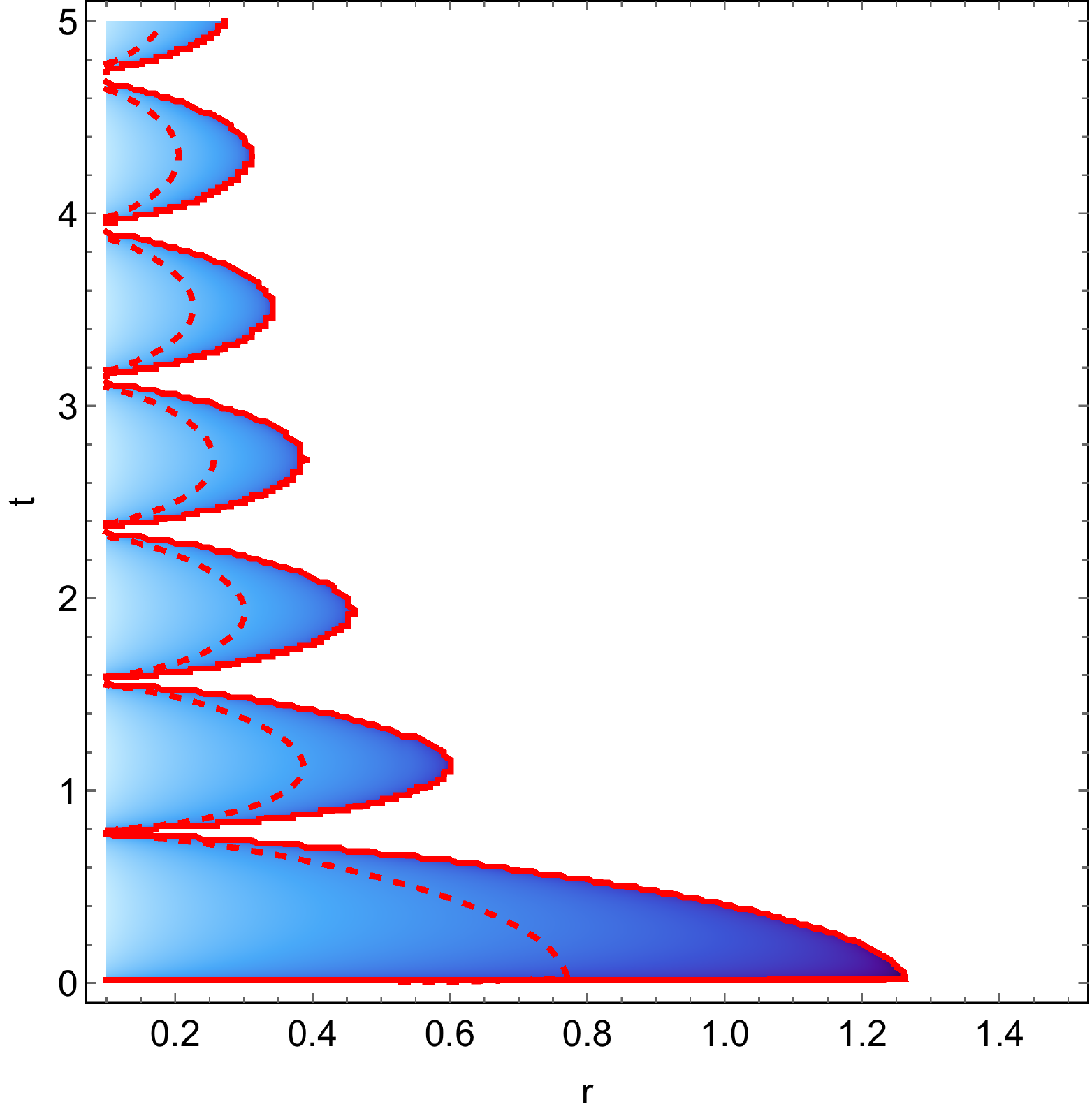}
   \includegraphics[width=0.3\linewidth,clip]{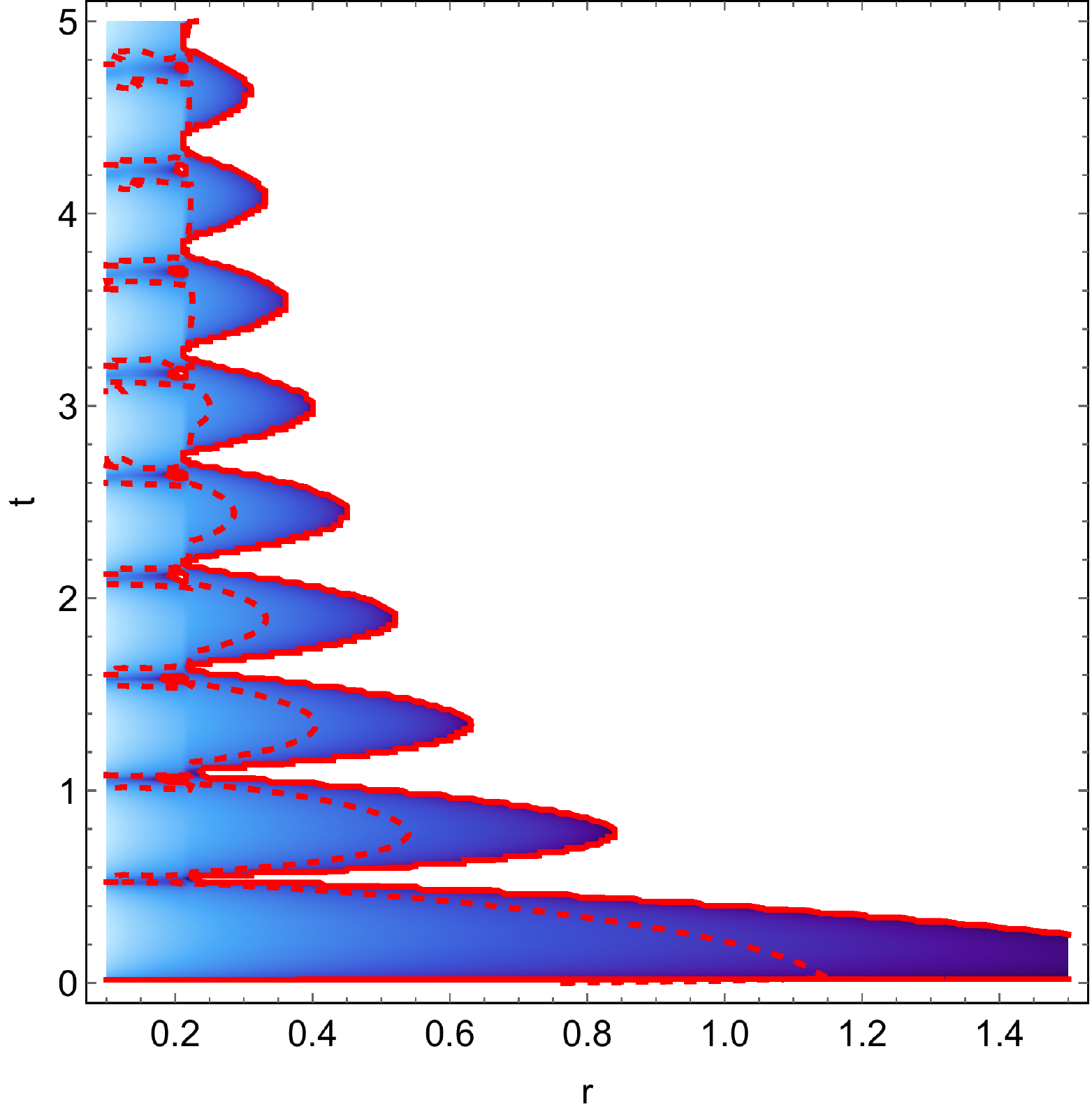}
   \caption{Evolution of negativity for $\omega=2, 4, 6$ with
     $\lambda=0.1, \sigma=0.4$. Detectors world lines are
     $r=$const. Detectors are entangled in regions enclosed by red
     solid lines (negativity zero lines). Violation of the Bell-CHSH
     inequality can be detected for parameters in regions enclosed by
     dotted lines. }
   \label{fig:mink-neg}
\end{figure}

We first show parameters for initial detection of
entanglement determined by Eq.~\eqref{eq:neg0} (Fig.~\ref{fig:mink-ent0}).
For any values of $r$, detection of entanglement is possible if we
choose  values of $\sigma$ above these lines.
\noindent
Our main interest is fate of detected entanglement after
evolution. Fig.~\ref{fig:mink-neg} contains  density plots of the negativity
in $(r,t)$ space and shows evolution of entanglement. The detector's
world line is $r=$constant in these figures. Although evolution of
negativity is different for different $r$ and $\omega$, if the
detectors do not catch the entanglement around $t\sim 0$, they cannot
detect non-zero negativity even after evolution.
\noindent
As examples of evolution of negativity, we show its time dependence
for $\omega=6,\sigma=0.4$ with two different $r$
(Fig.~\ref{fig:mink-negt}):
\begin{figure}[H]
    \centering
     \includegraphics[width=0.38\linewidth,clip]{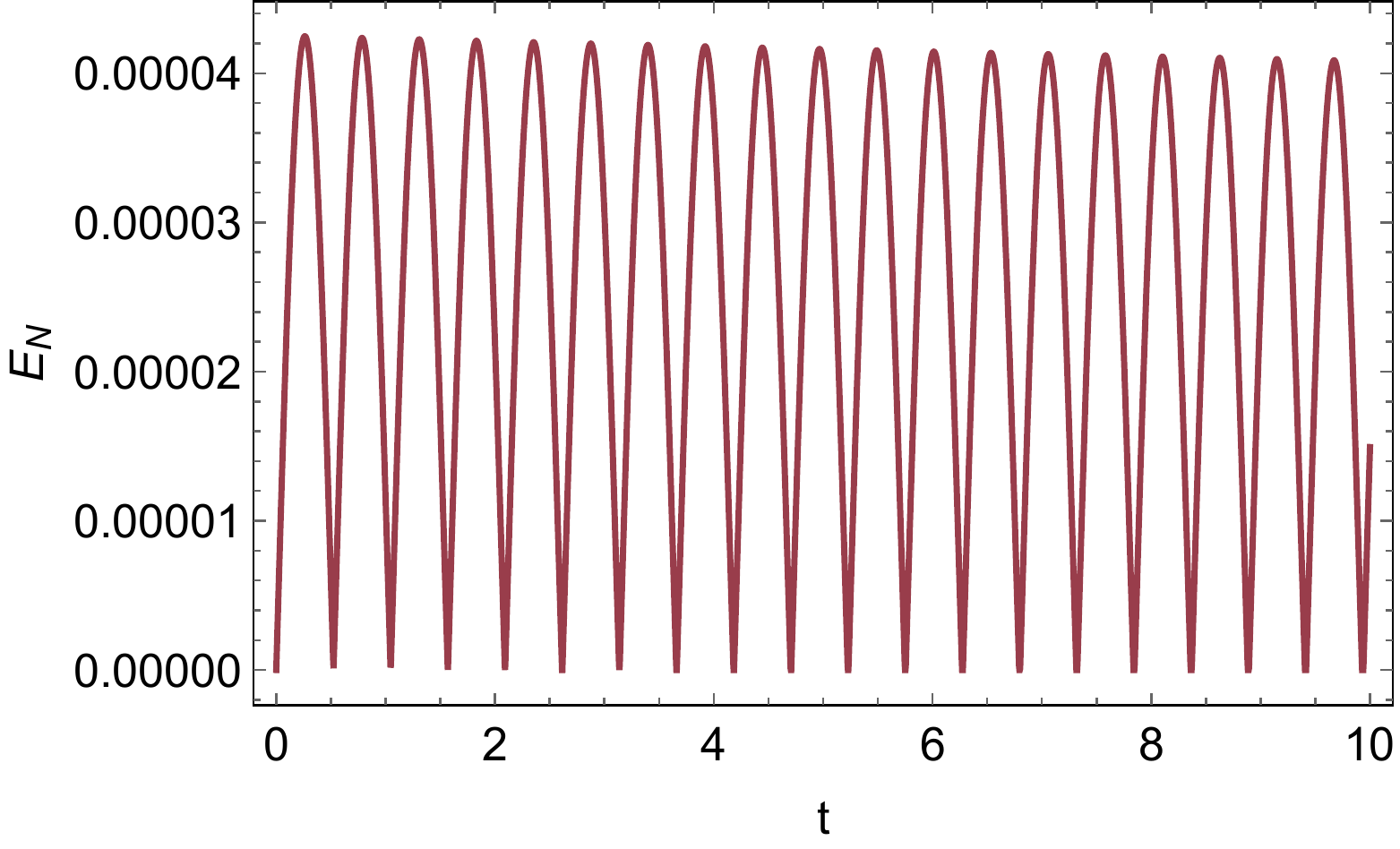}
    \includegraphics[width=0.4\linewidth,clip]{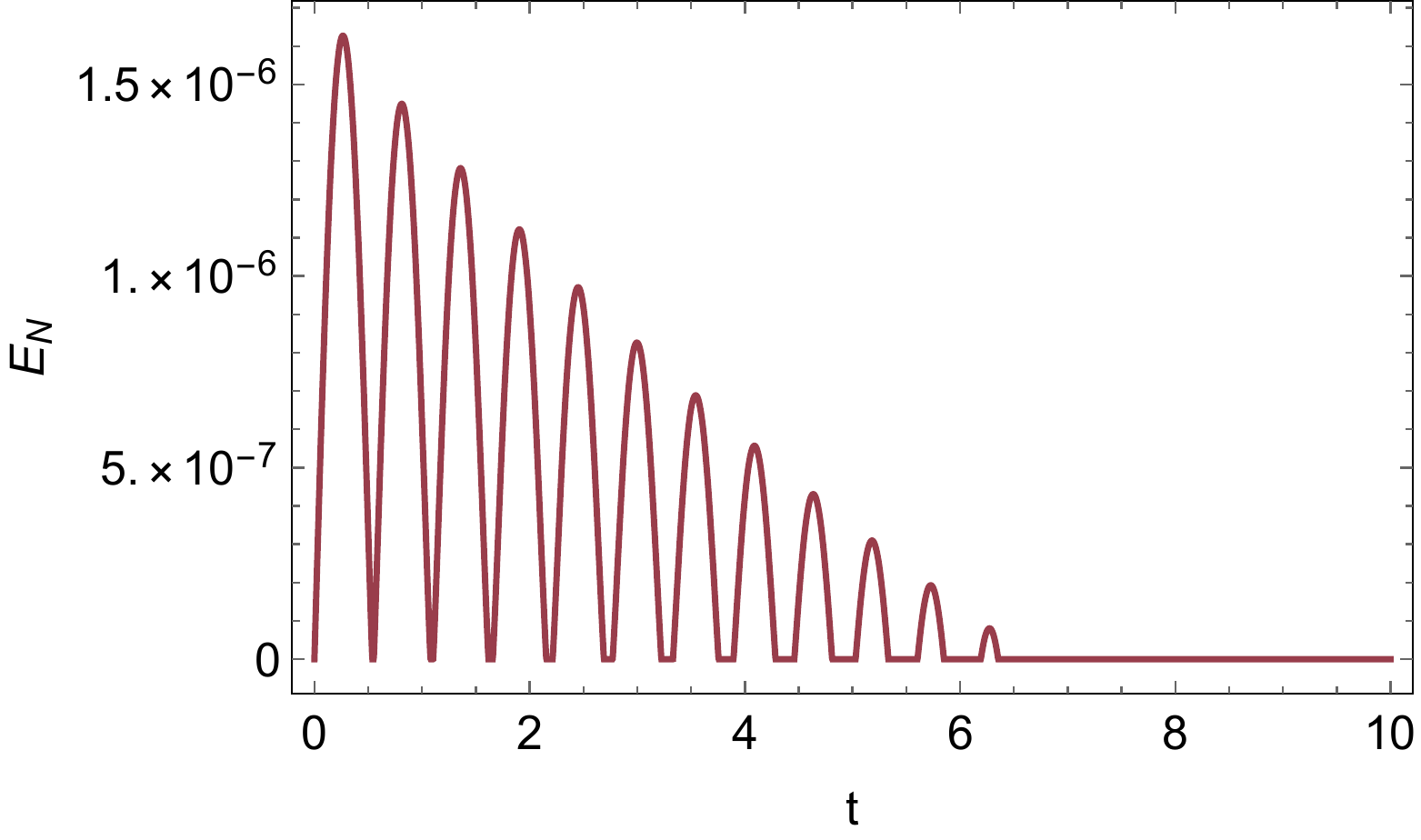}
    \caption{Time dependence of the negativity for
      $\lambda=0.1, \omega=6, \sigma=0.4$. Left: $r=0.05$. Right: $r=0.25$.}
    \label{fig:mink-negt}
\end{figure}
\noindent
The negativity shows oscillatory behavior with a period $2\pi/\omega$.
For both $r$, detectors can catch the entanglement of the field
initially. For $r=0.05$, after initial detection of entanglement, the
negativity decays linearly in time with oscillation. For $r=0.25$, the
negativity decays also linearly in time, and after death and revival of
entanglement several times, the negativity settles down to zero. 
Concerning the Bell-CHSH inequality, as shown in
Fig.~\ref{fig:mink-neg}, violation of the inequality can be detectable
for parameters in regions enclosed by dotted lines. These regions are
contained in regions determined by zero negativity lines because
non-zero values of the negativity provide a necessary condition for
violation of the Bell-CHSH inequality.

\begin{figure}[H]
    \centering
    \includegraphics[width=0.3\linewidth,clip]{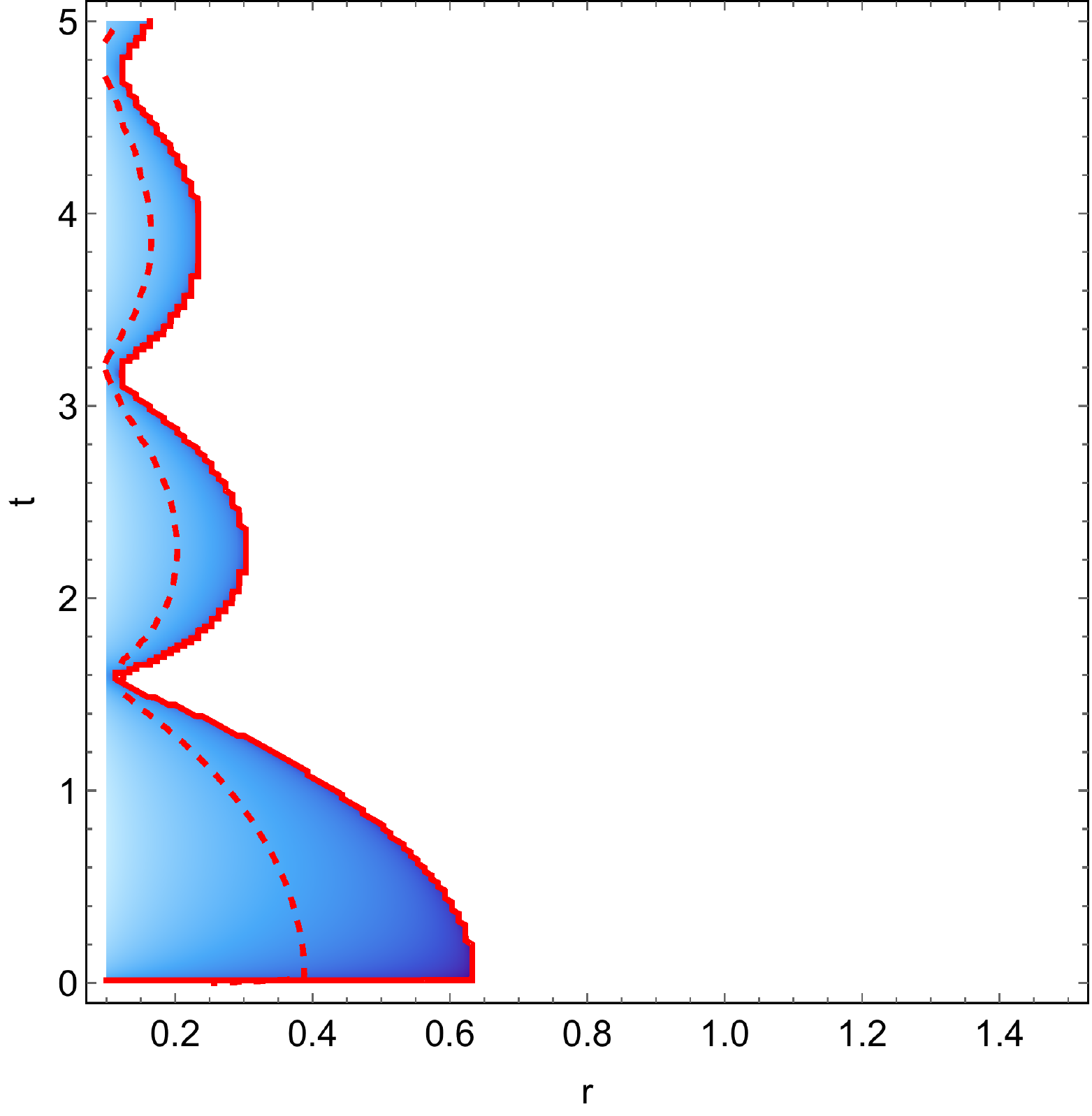}
    \includegraphics[width=0.3\linewidth,clip]{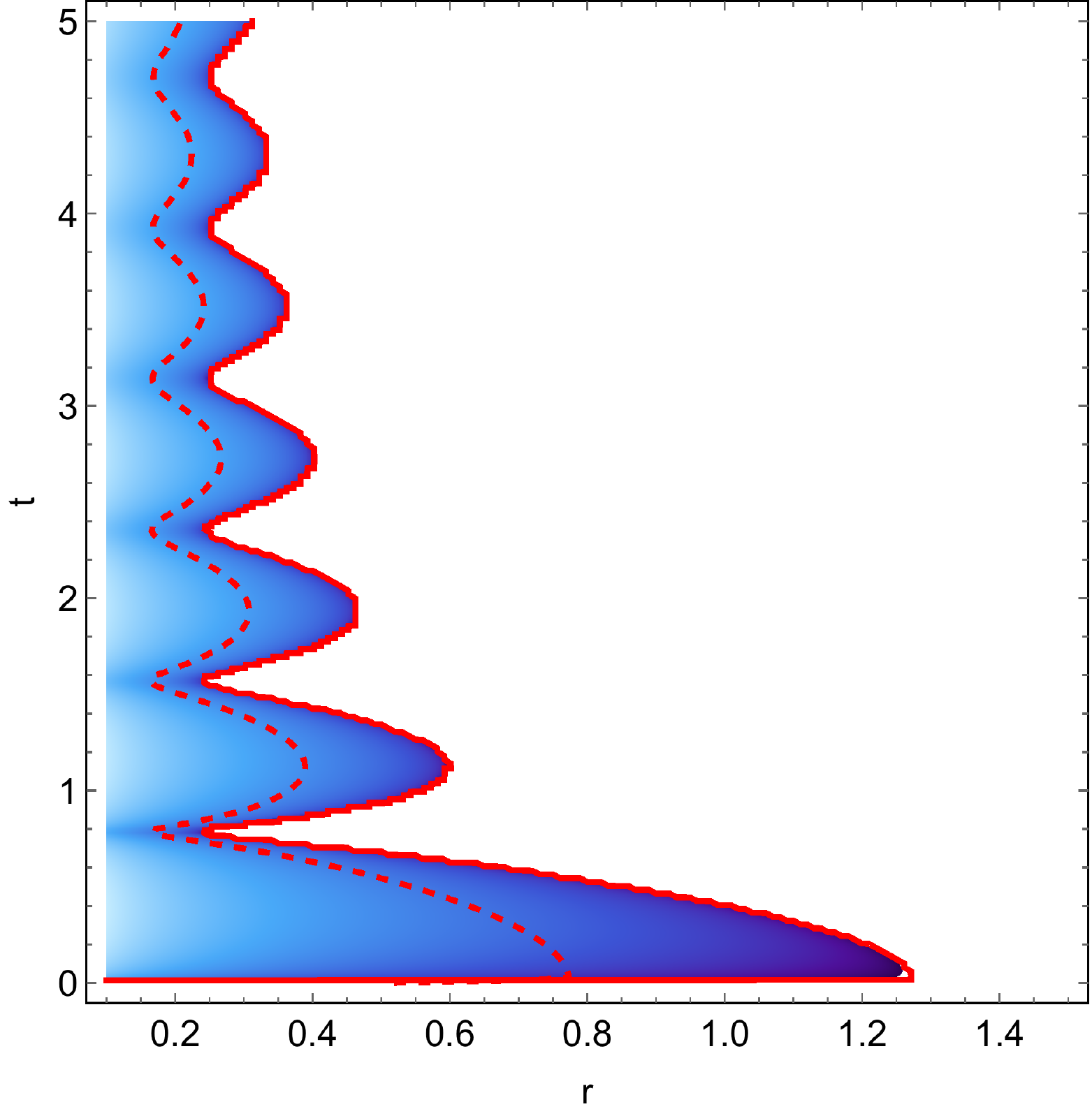}
   \includegraphics[width=0.3\linewidth,clip]{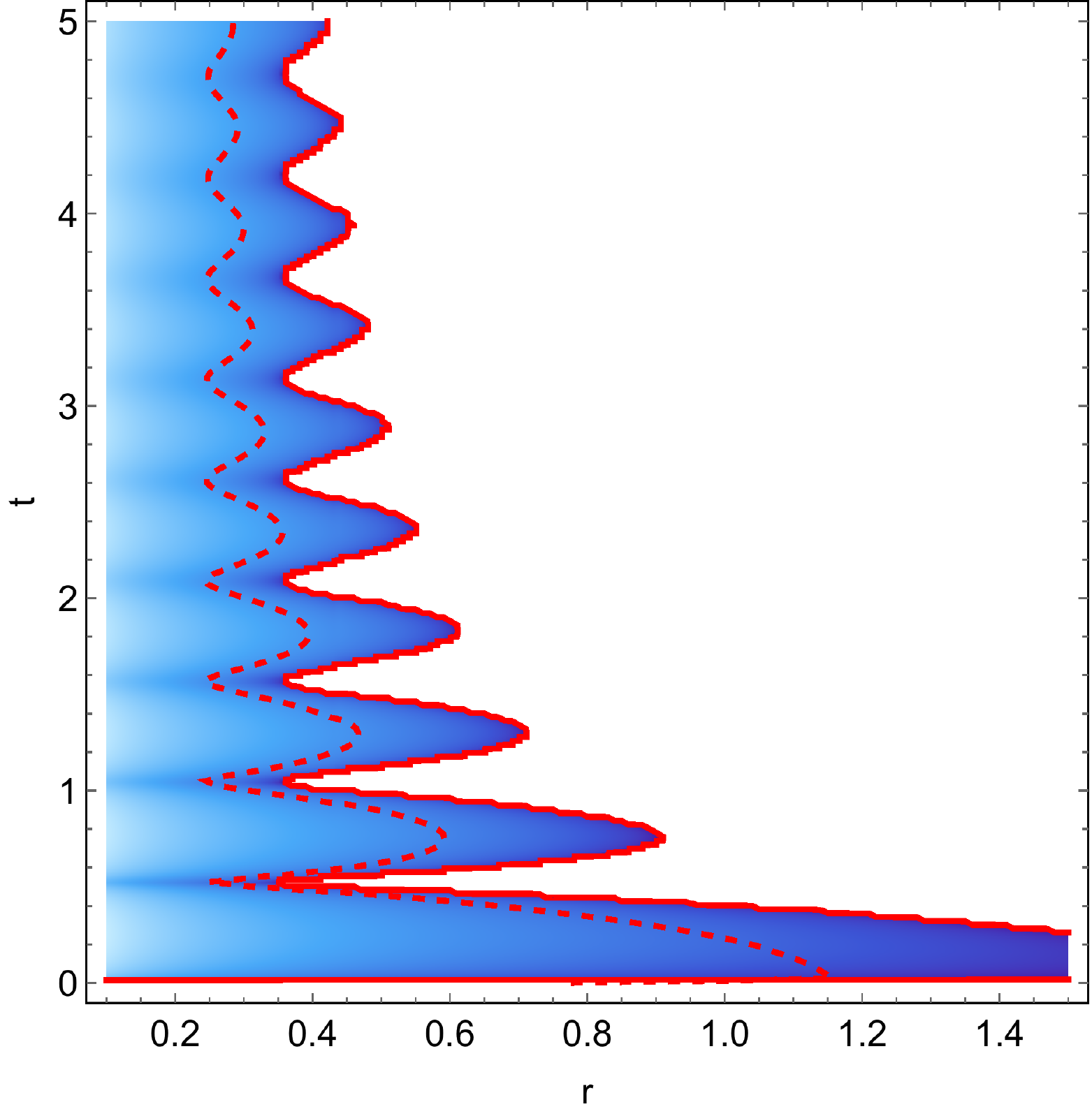}
   \caption{Evolution of negativity for $\omega=2, 4, 6$ with
     $\lambda=0.5, \sigma=0.4$. Detectors world lines are
     $r=$const. Detectors are entangled in regions enclosed by red
     solid lines (negativity zero lines). Violation of the Bell-CHSH
     inequality can be detected for parameters in regions enclosed by
     dotted lines. }
   \label{fig:mink-negb}
\end{figure}

\begin{figure}[H]
    \centering
    \includegraphics[width=0.38\linewidth,clip]{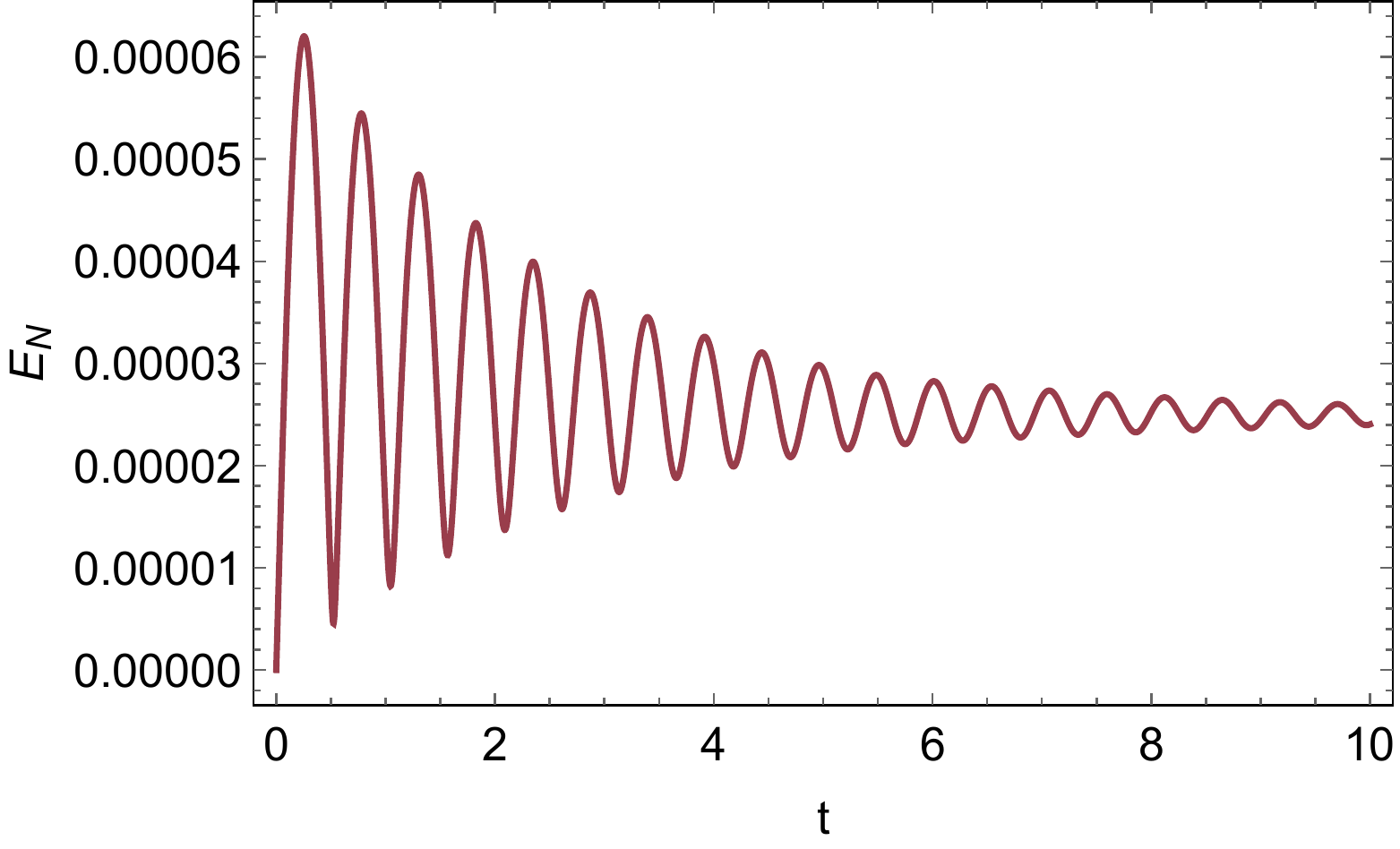}
    \includegraphics[width=0.4\linewidth,clip]{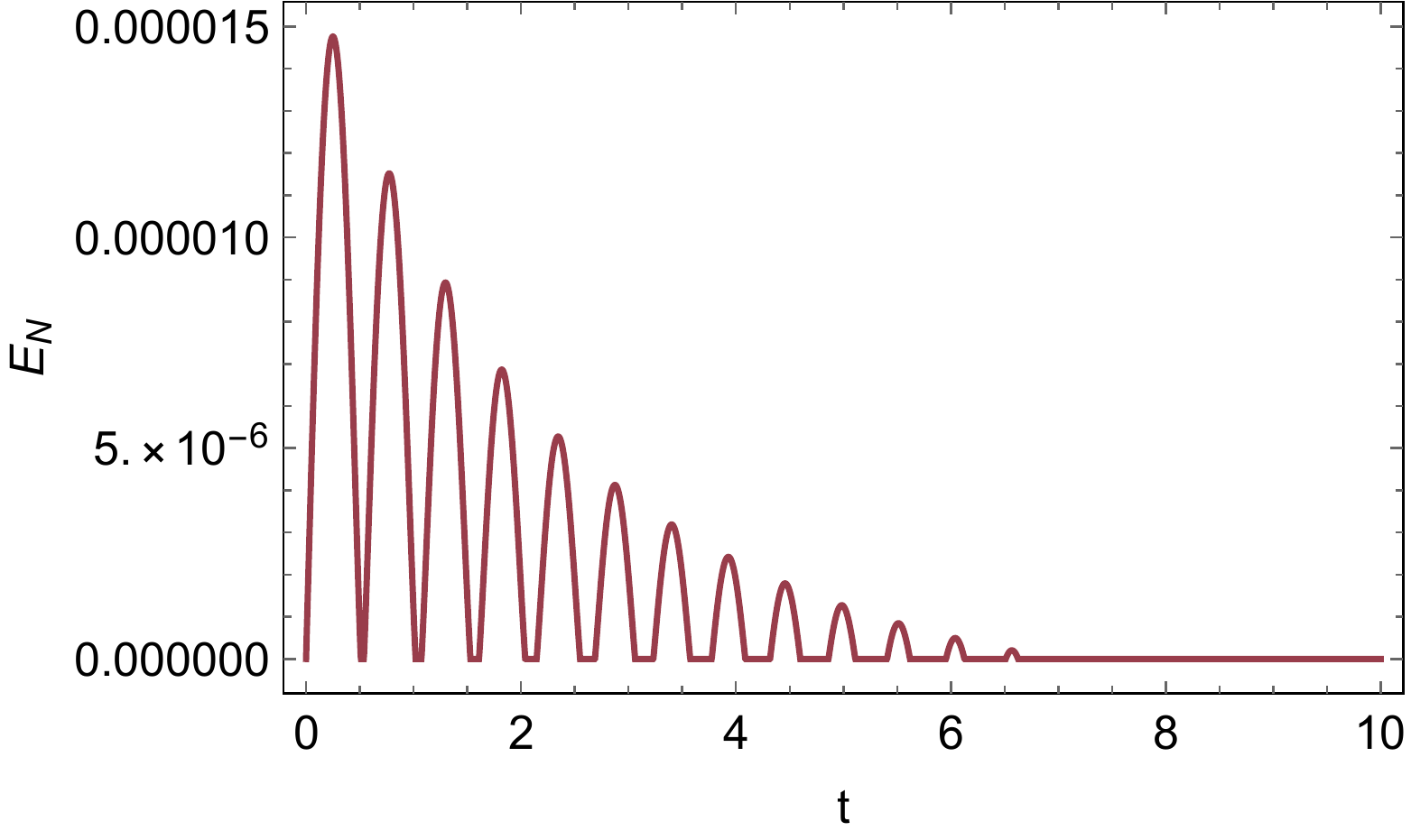}
    \caption{Time dependence of the negativity for
      $\lambda=0.5, \omega=6, \sigma=0.4$. Left: $r=0.2$. Right: $r=0.4$.}
    \label{fig:mink-negtb}
\end{figure}

For a larger value of the coupling constant $\lambda=0.5$, the
negativity evolves as follows (Fig.~\ref{fig:mink-negb},
Fig.~\ref{fig:mink-negtb})
\noindent
For $r=0.2$, after initial detection of entanglement, the negativity
approaches a non-zero constant value. But for $r=0.4$, after death and
revival of entanglement several times, the negativity settles down to
zero and the state of the detectors finally becomes separable. From
Fig.~\ref{fig:mink-negb}, we can observe that the detectors' system
approaches stationary state after elapsed time $\sim \sigma/\lambda^2$.

\subsection{Massless scalar field in de Sitter space}
We consider the massless conformal scalar field and the massless minimal
scalar field in de Sitter space.  

\subsubsection{Conformal scalar}
\begin{figure}[H]
  \centering
   \includegraphics[width=0.4\linewidth,clip]{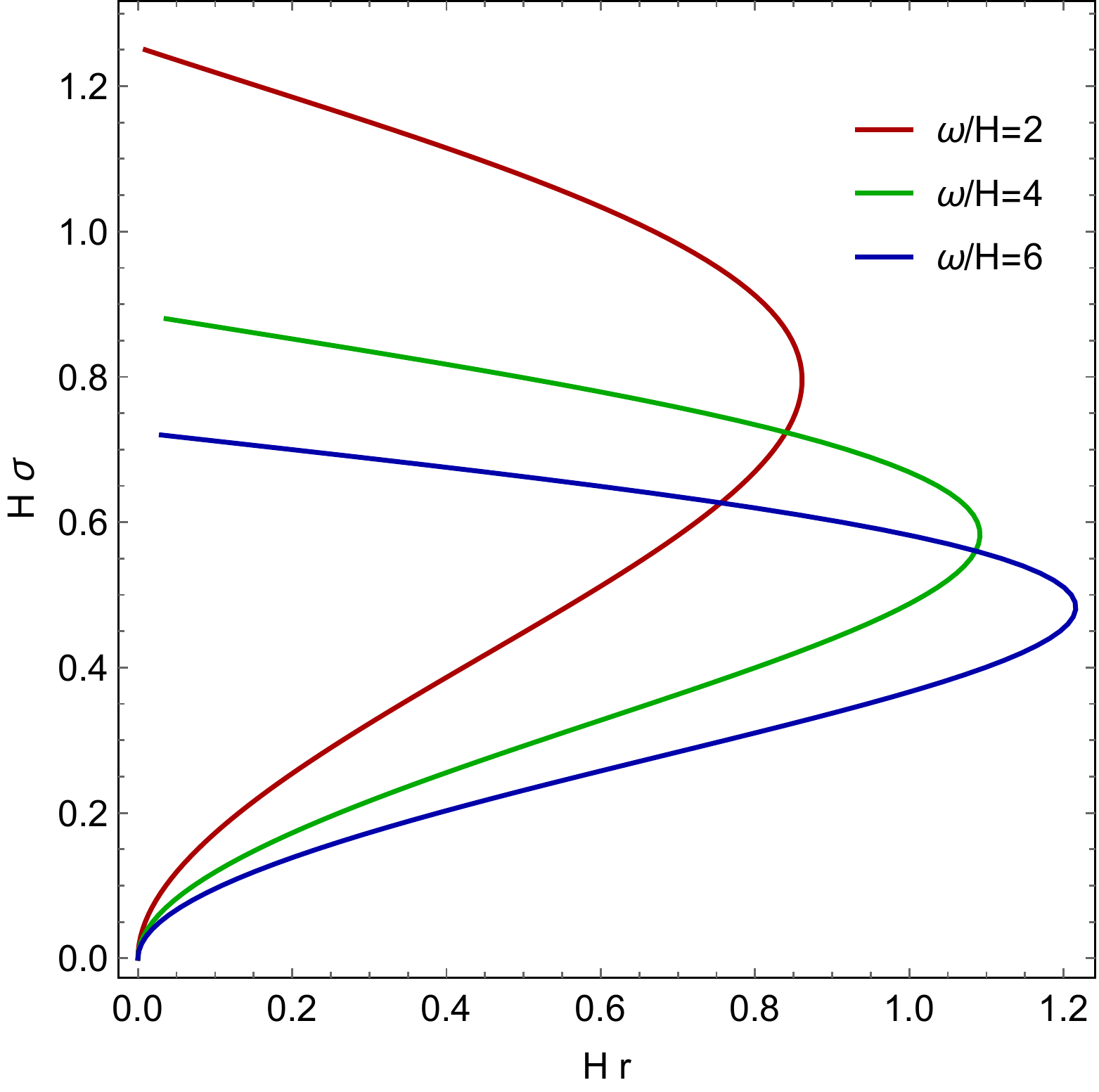}
   \caption{Parameters of entanglement detection around $t=0$. 
     Entanglement can be detected for parameters in
     regions enclosed by each lines.}
  \label{fig:neg0-conf}
\end{figure}

For the massless conformal scalar field, the negativity around initial
time is determined by Eq.~\eqref{eq:neg0}
\begin{equation}
  |c_{11}^1|-c^0_{21}=\frac{\lambda^2e^{-\omega^2\sigma^2}}{2\pi^2\sigma}(\sigma
  H)^2\left[
    \frac{e^{-2H\sigma}}{(Hr)^2}-\frac{1}{4\sin^2(H\omega\sigma^2)}\right].
\end{equation}
From this, we can estimate the maximum distance for entanglement
detection (Fig.~\ref{fig:neg0-conf}) 
\begin{equation}
  r\le\frac{2}{H}\,e^{-\left(\frac{\pi H}{2\omega}\right)^{1/2}}<\frac{2}{H}.
\end{equation}
\noindent
Thus,  for large values of $\omega$,
initial detection of entanglement beyond the Hubble horizon scale up
to $2H^{-1}$ is possible for the conformal scalar field.

\begin{figure}[H] 
  \centering
   \includegraphics[width=0.3\linewidth,clip]{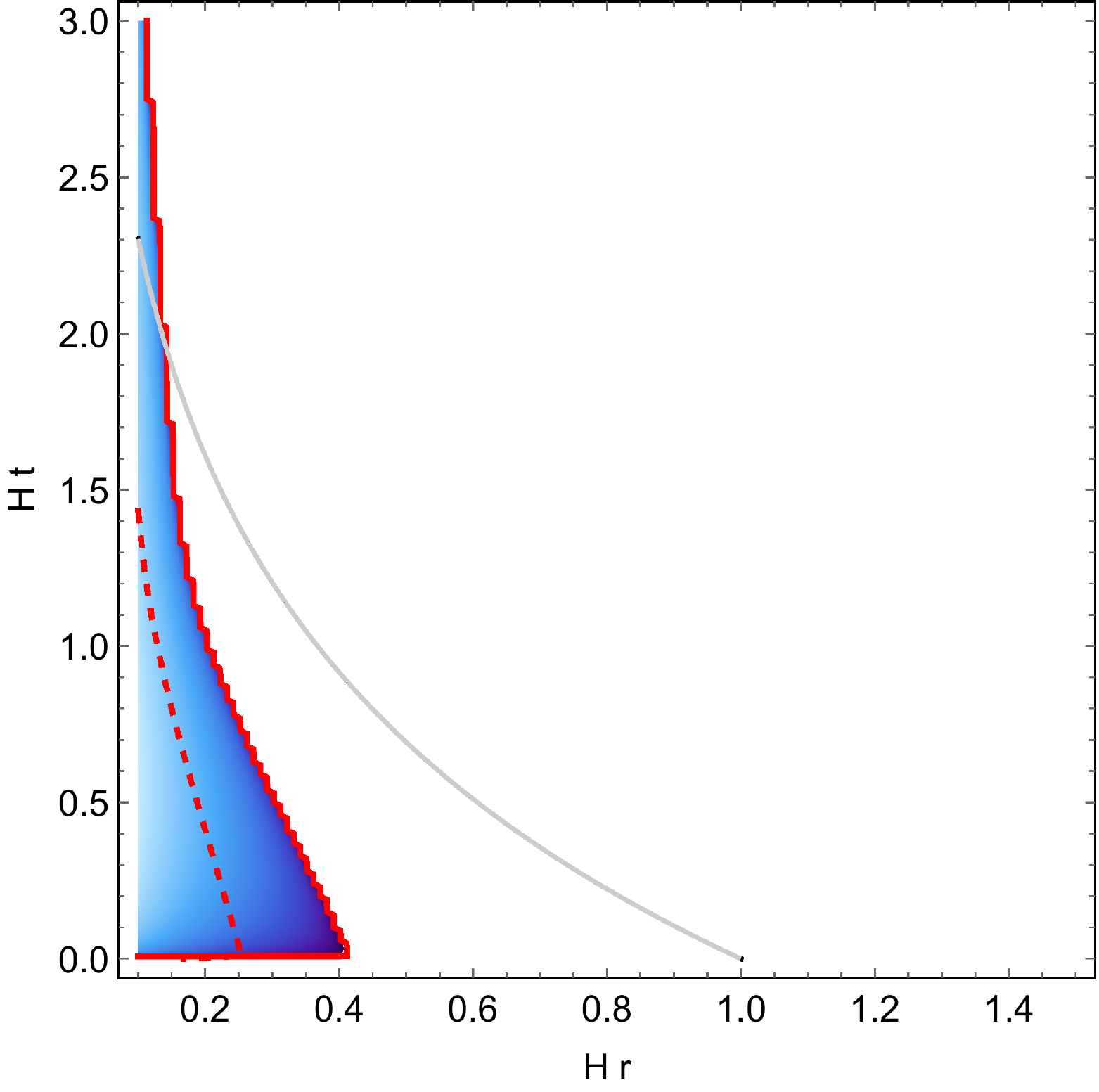}
   \includegraphics[width=0.3\linewidth,clip]{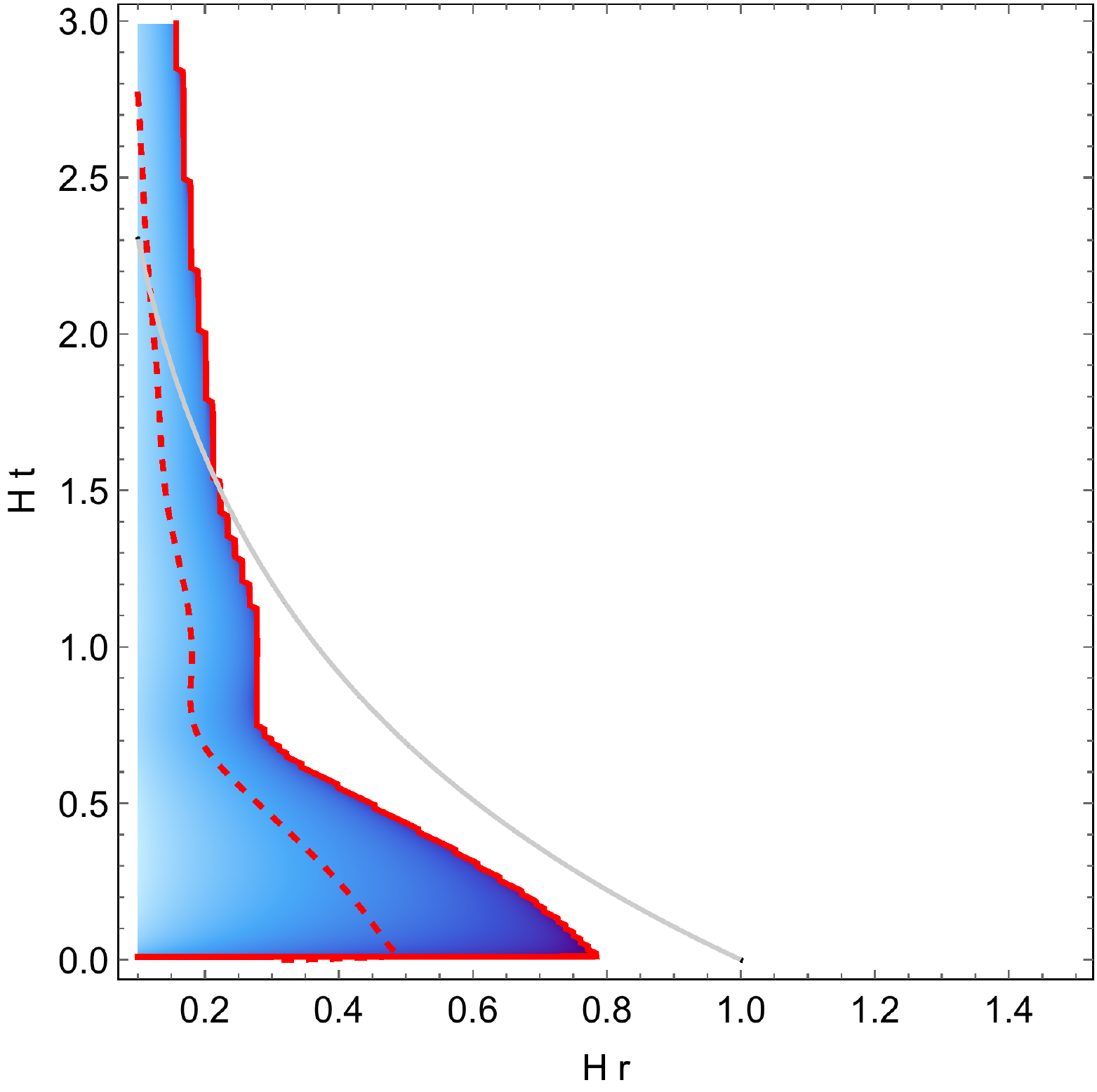}
  \includegraphics[width=0.3\linewidth,clip]{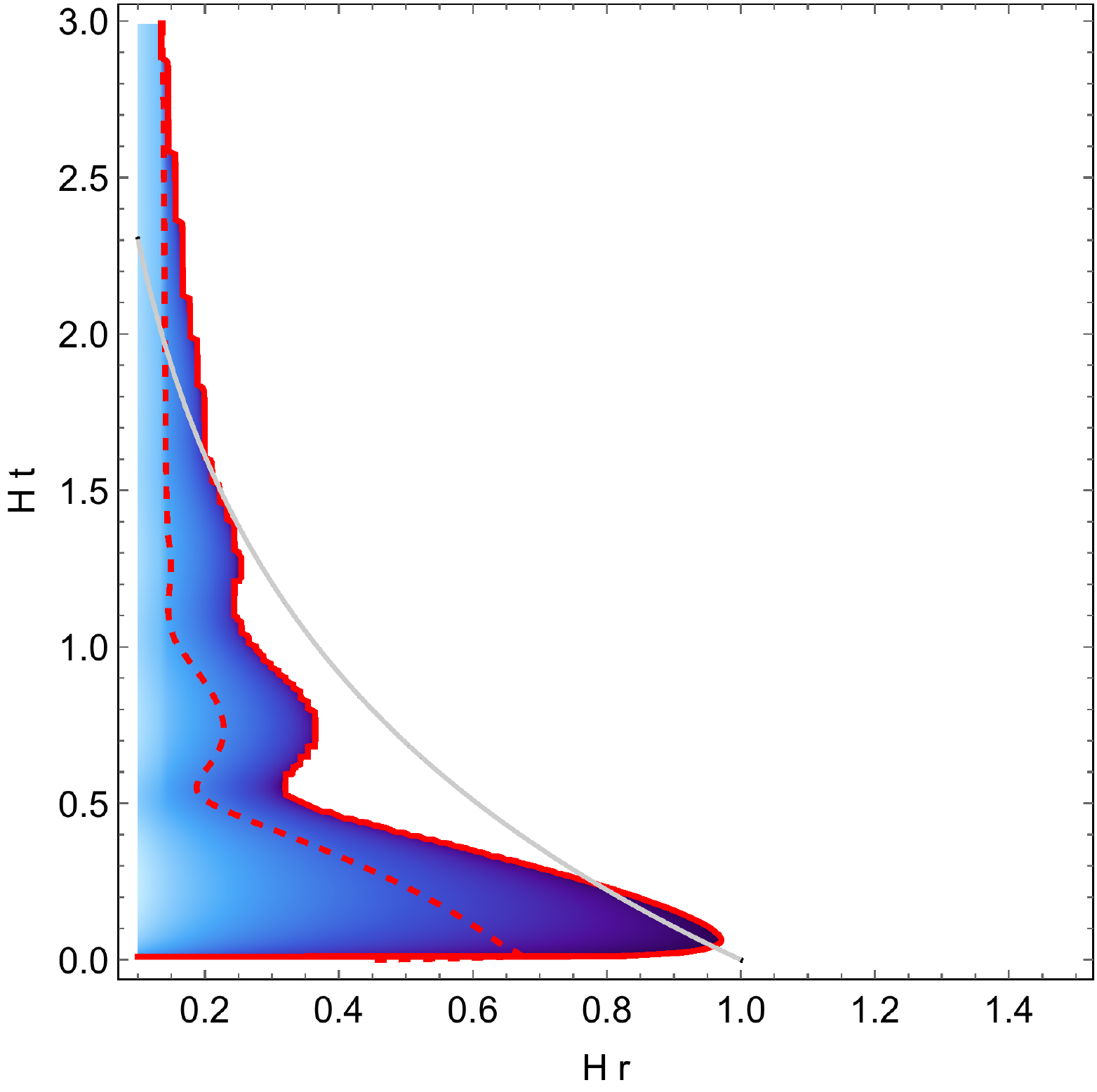}
  \caption{Evolution of negativity between comoving detectors for
    $\omega/H=2,4,6$ with $\lambda=0.1,\sigma H=0.4$. The detector's
    world lines are $r=\text{const.}$ The gray solid lines represent
    comoving size of the Hubble horizon. The red solid lines represent
    negativity zero contours.  Detectors are entangled for parameters
    in regions enclosed by the solid red lines. For parameters in
    regions enclosed by red dotted lines, violation of Bell's
    inequality can be detected. }
\label{fig:neg-conf}
\end{figure}

Evolution of the negativity is shown in Fig.~\ref{fig:neg-conf}.
After initial detection of entanglement, the negativity finally
becomes zero for large separation. For $\omega/H=2,4$, initial
detection of entanglement is possible only for $r<H^{-1}$ and the
detected entanglement survives beyond the super horizon scale
$>H^{-1}$ if the separation of two detectors is small enough.  For
$\omega/H=6$, initial detection of entanglement for super horizon
scale is possible (about $1.1H^{-1}$ in this case). Also in this case,
for sufficiently small $r$, the detected entanglement survives when
the physical separation of two detectors exceeds the Hubble horizon
scale.  Violation of Bell-CHSH inequality beyond the Hubble horizon
scale detectable for sufficiently small $r$ (middle and right panels
of Fig.~\ref{fig:neg-conf}).

\begin{figure}[H]
  \centering
  \includegraphics[width=0.4\linewidth,clip]{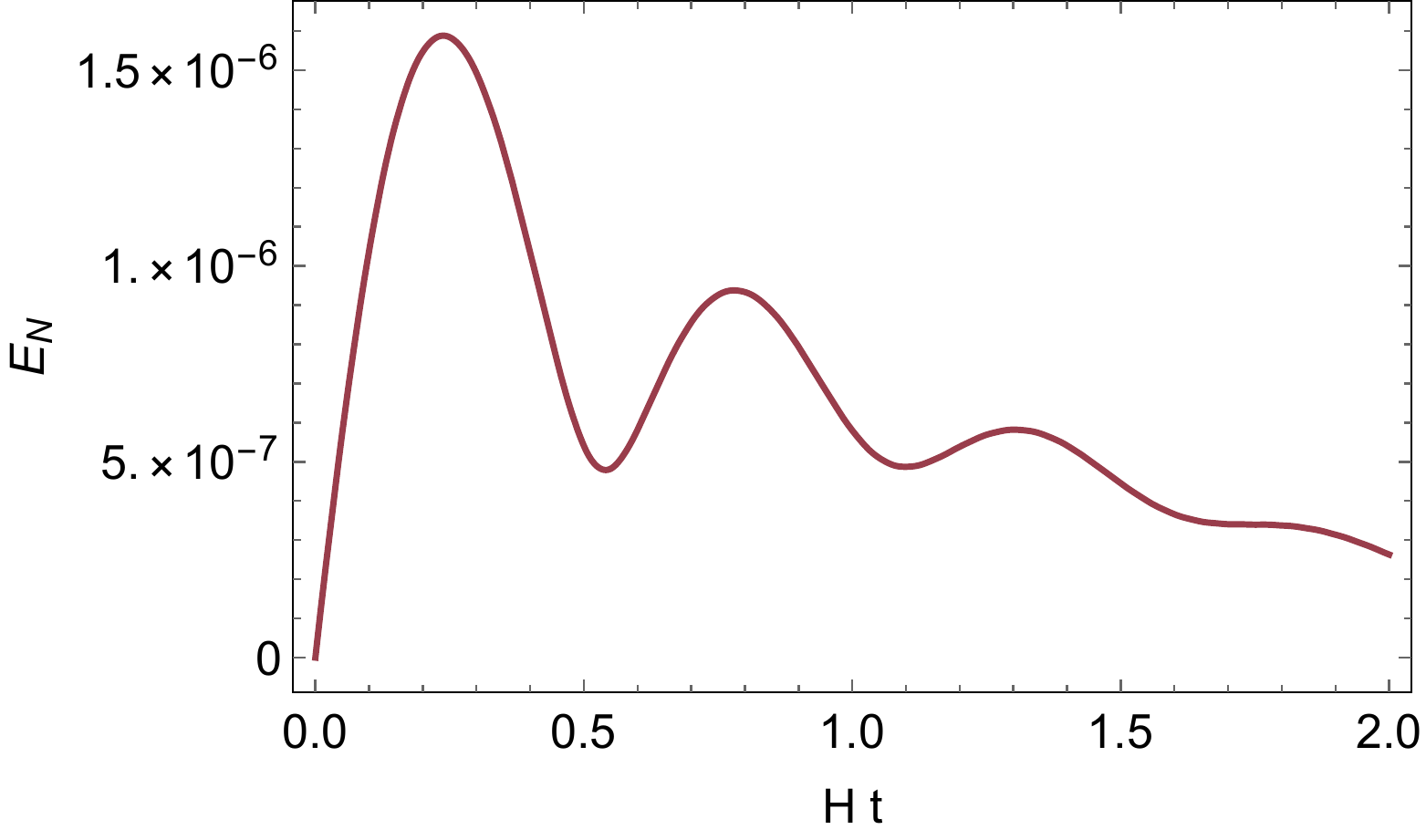}
  \includegraphics[width=0.4\linewidth,clip]{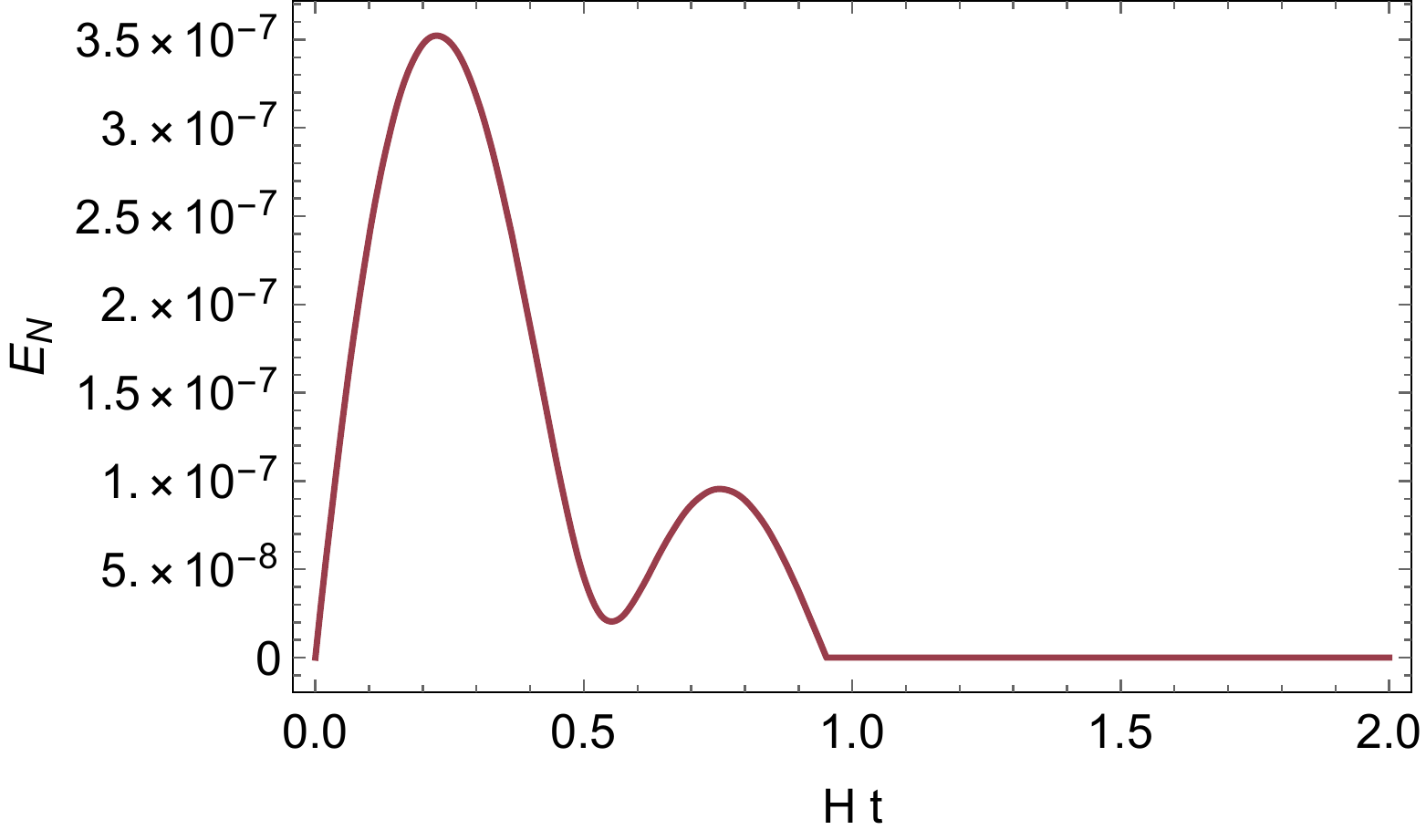}
  \caption{Evolution of negativity for $\lambda=0.1,\omega/H=6$ at $rH=0.15$ (left)
     and at $rH=0.3$ (right).}
\label{fig:conf-negt}
\end{figure}

\noindent
Fig.~\ref{fig:conf-negt} shows time evolution of the negativity for
$\omega/H=6$ with two different $r$. Behavior of evolution depends on
$r$. For $r H=0.15$, the negativity remains non-zero after horizon
exit. For $rH=0.3$, entanglement finally becomes zero.

\begin{figure}[H] 
  \centering
   \includegraphics[width=0.3\linewidth,clip]{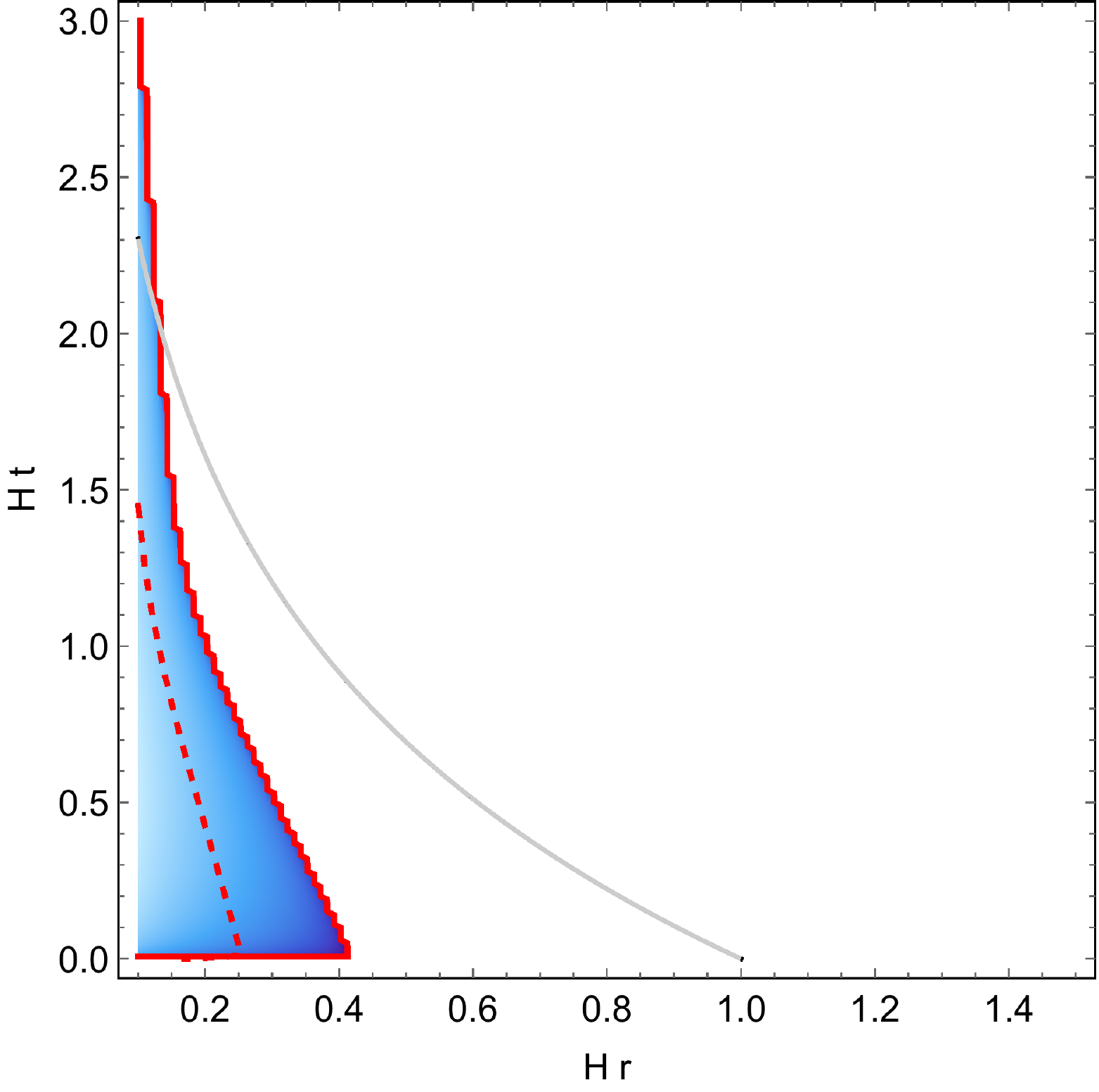}
   \includegraphics[width=0.3\linewidth,clip]{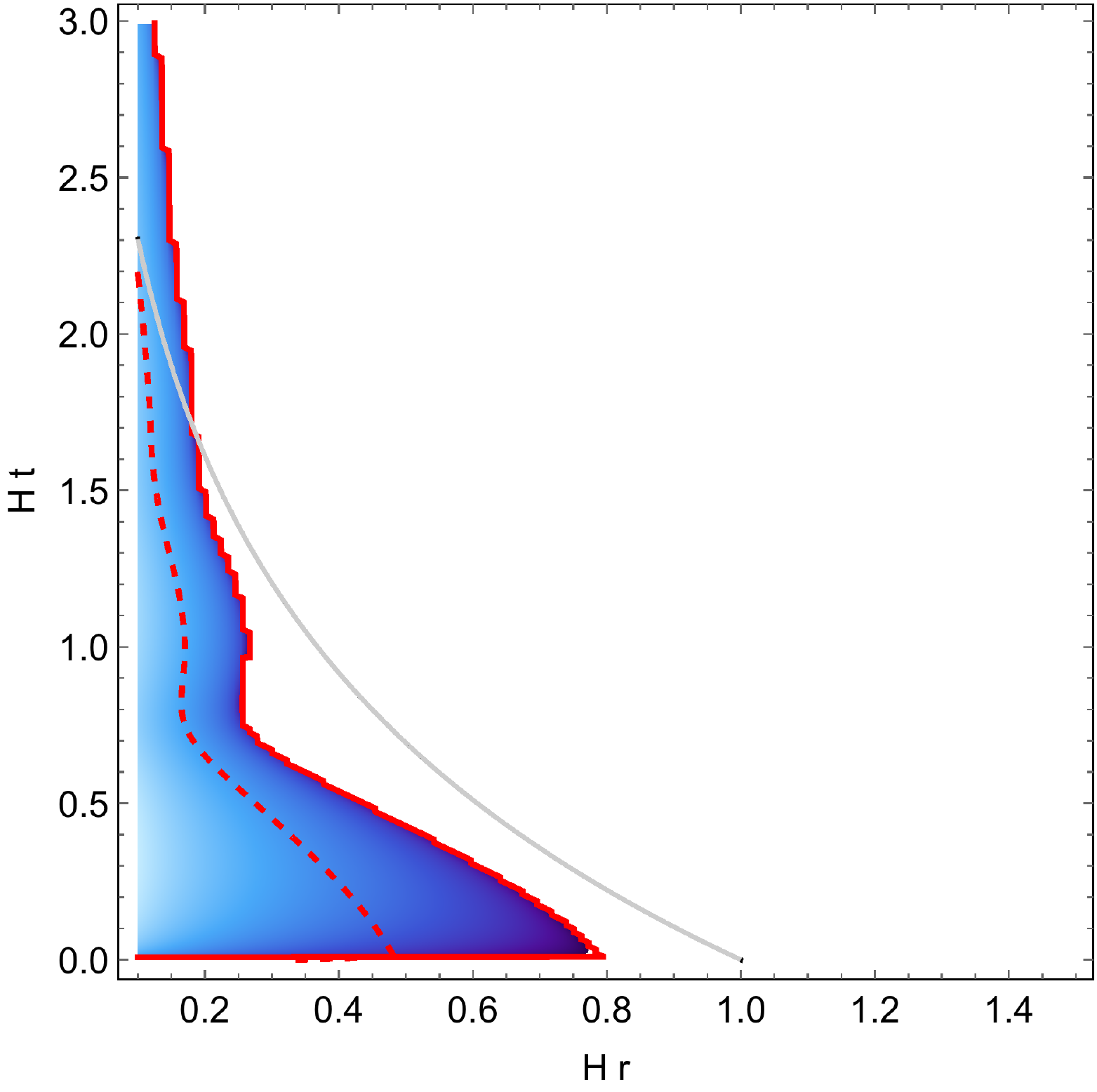}
  \includegraphics[width=0.3\linewidth,clip]{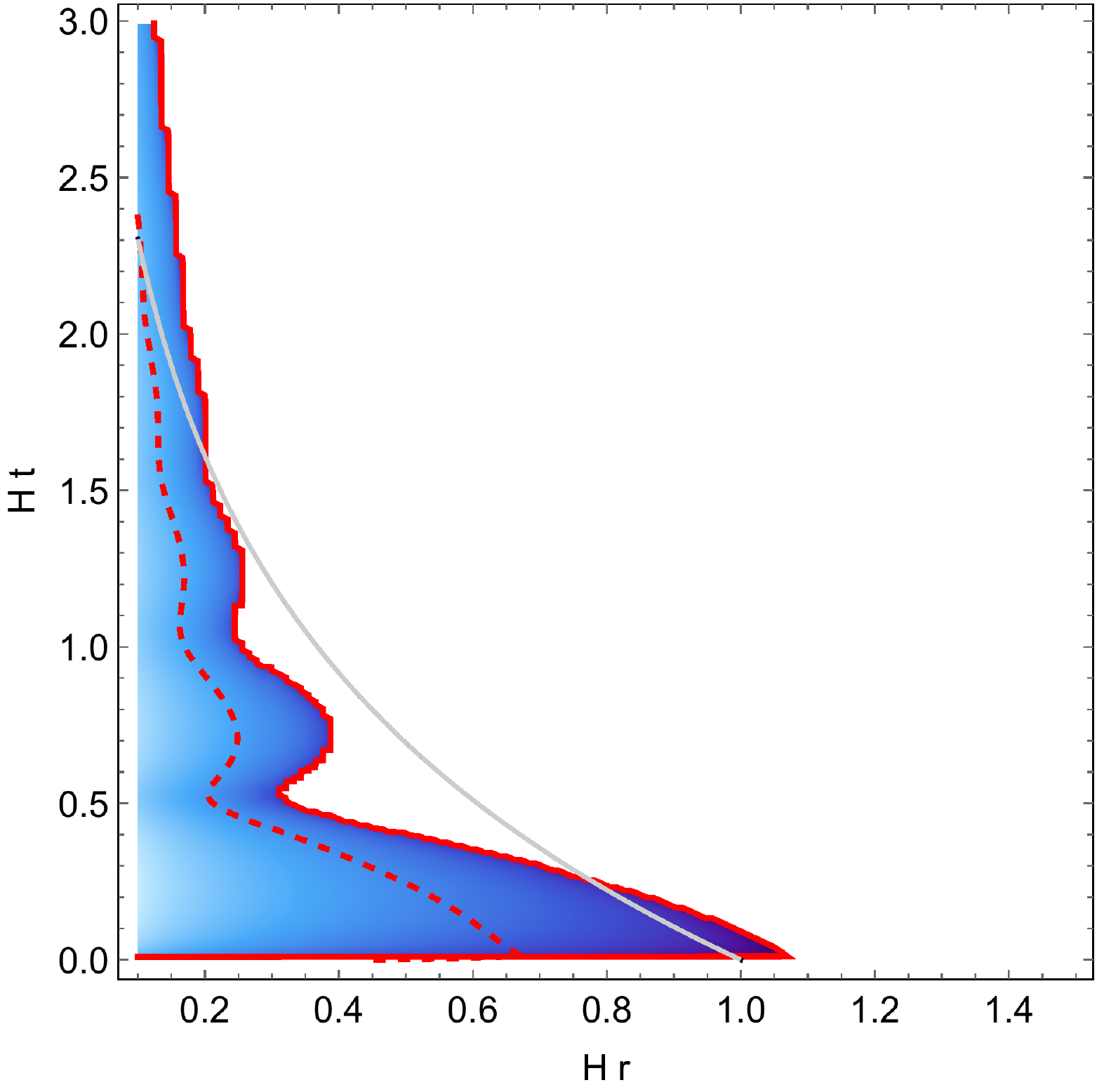}
  \caption{Evolution of negativity between comoving detectors for
    $\omega/H=2,4,6$ with $\lambda=0.5,\sigma H=0.4$. The detector's
    world lines are $r=\text{const.}$ The gray solid lines represent
    comoving size of the Hubble horizon. The red solid lines represent
    negativity zero contours.  Detectors are entangled for parameters
    in regions enclosed by the solid red lines. For parameters in
    regions enclosed by red dotted lines, violation of Bell's
    inequality can be detected. }
\label{fig:neg-confb}
\end{figure}

\begin{figure}[H]
  \centering
  \includegraphics[width=0.4\linewidth,clip]{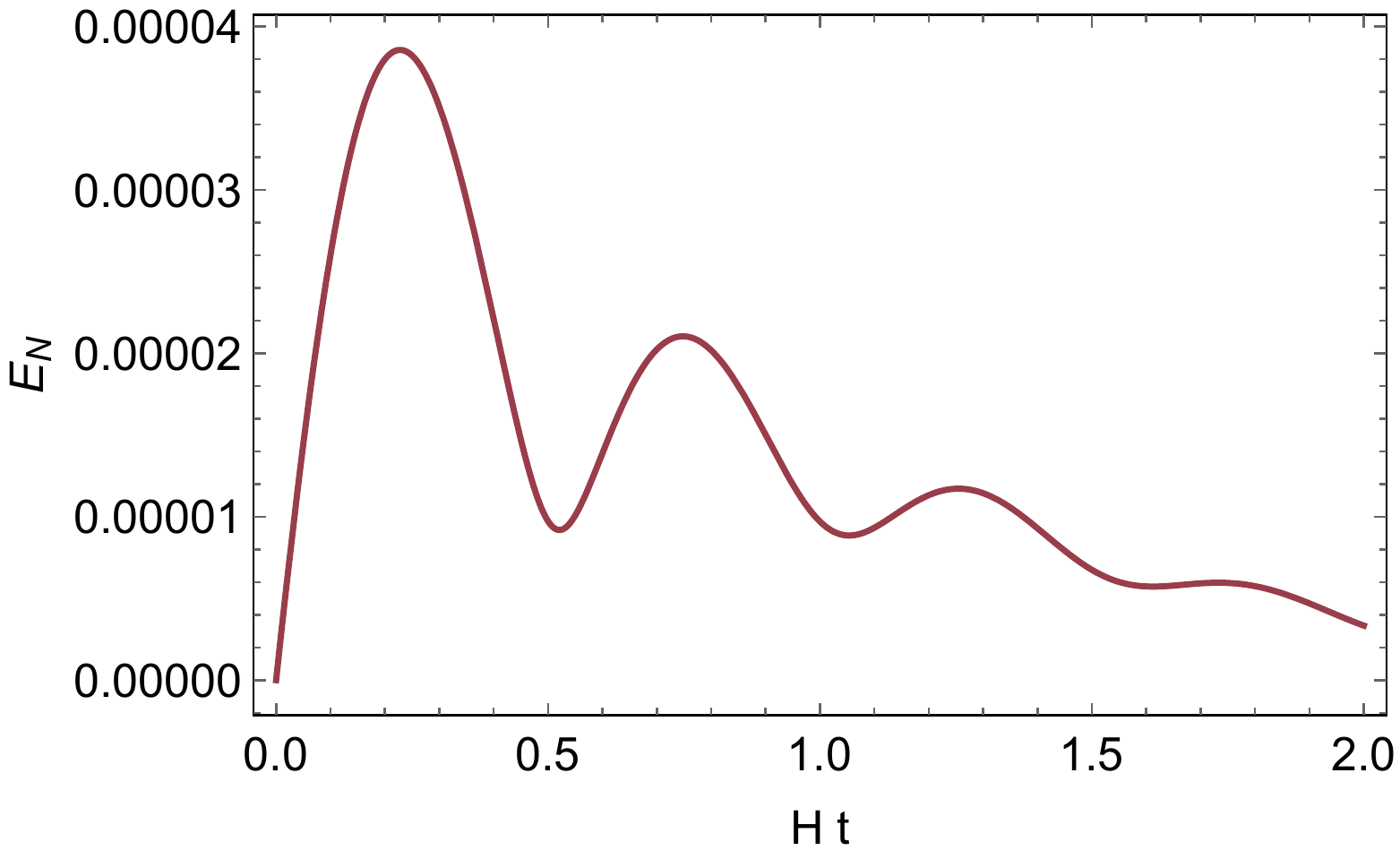}
  \includegraphics[width=0.4\linewidth,clip]{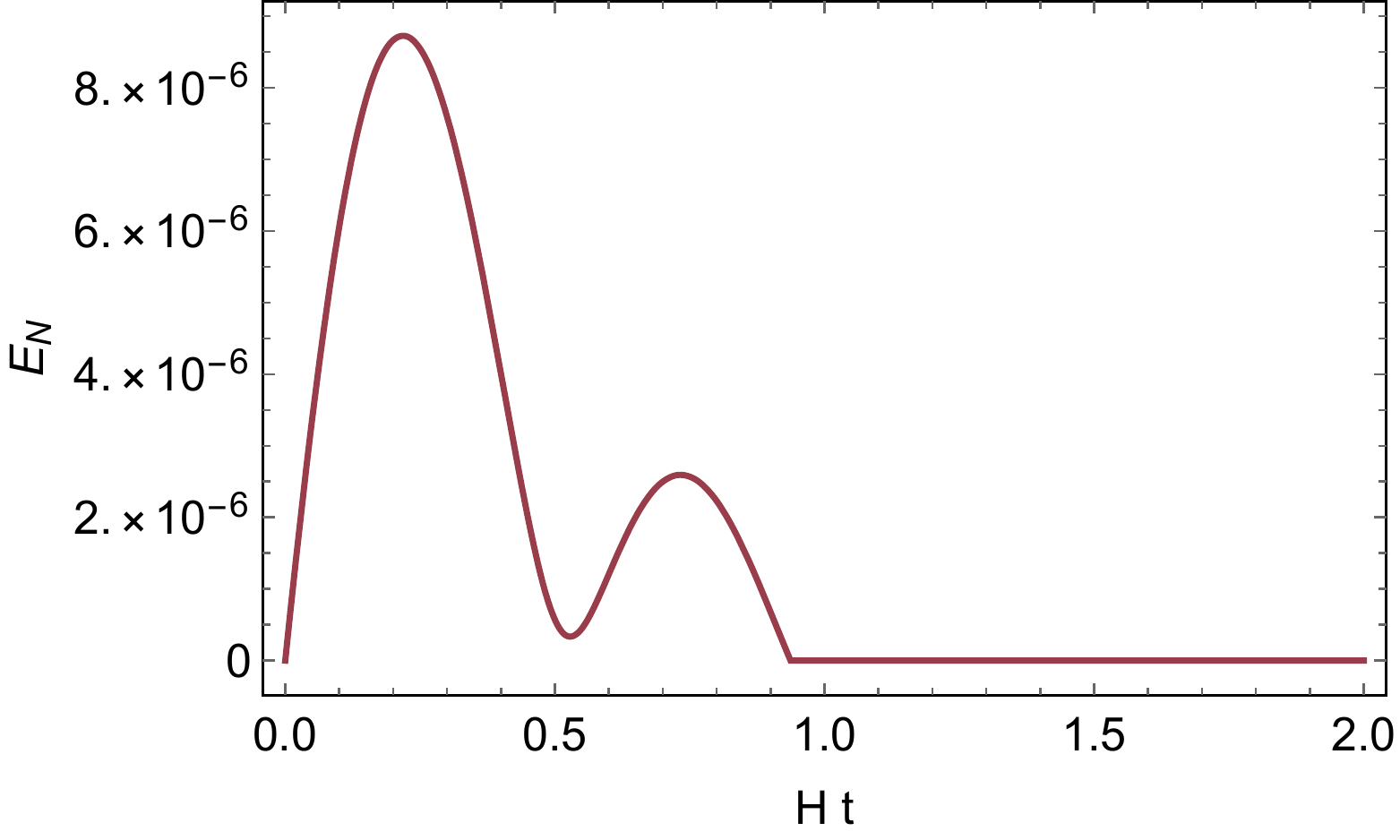}
  \caption{Evolution of negativity for $\lambda=0.5,\omega/H=6$ at $rH=0.15$ (left)
     and at $rH=0.3$ (right).}
\label{fig:conf-negtb}
\end{figure}

Evolution of negativity for $\lambda=0.5$ is  shown in
Fig.~\ref{fig:neg-confb} and Fig.~\ref{fig:conf-negtb}. Also in this
case, detected entanglement  can survive beyond the
Hubble horizon scale for small $r$.

\subsubsection{Massless minimal scalar}

\begin{figure}[H]
  \centering
  \includegraphics[width=0.4\linewidth,clip]{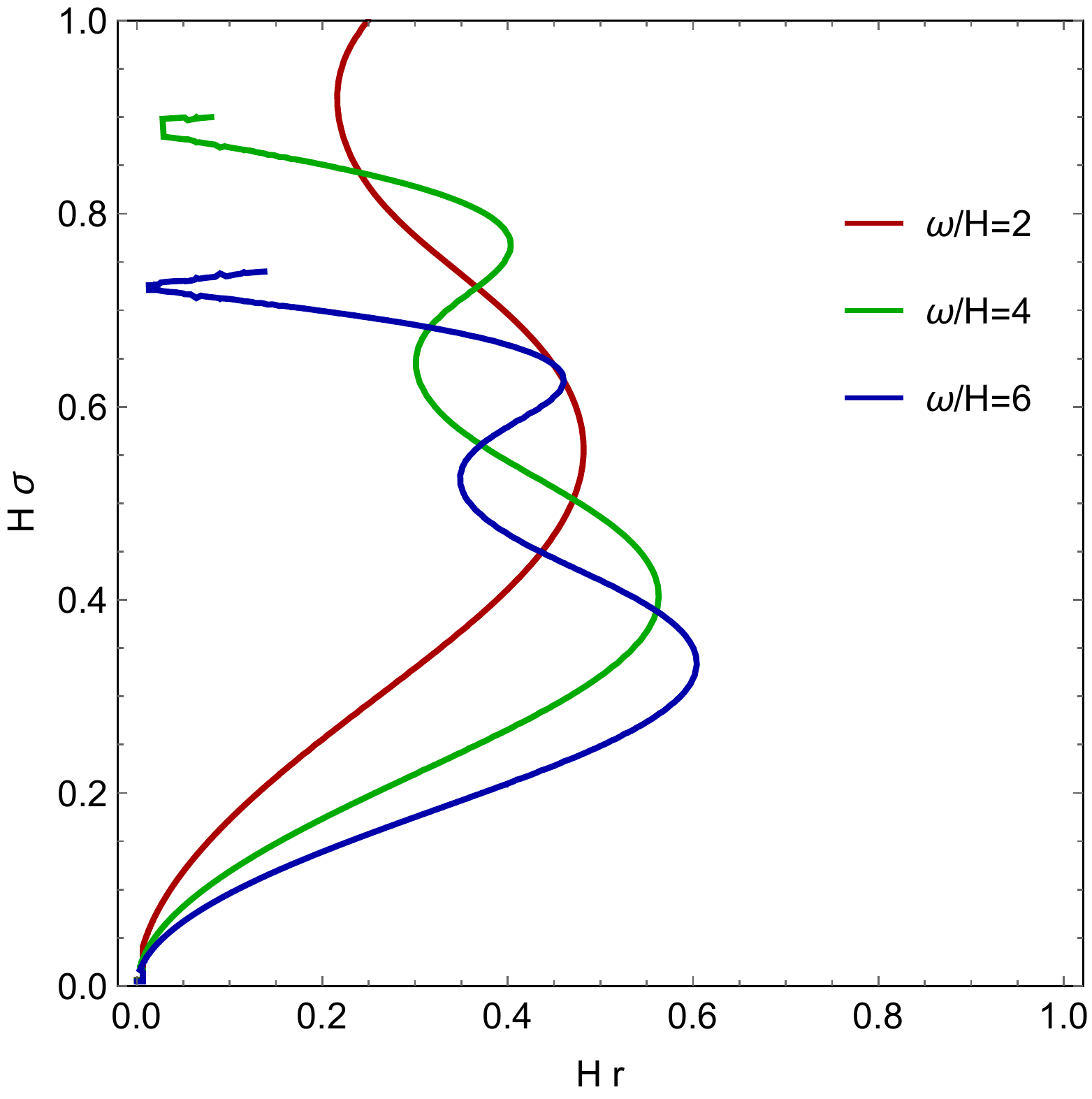}
  \caption{Parameters of initial detection of
    entanglement. Entanglement of the minimal scalar field can be
    detected for parameters in the regions enclosed by each lines.}
\label{fig:mini-neg0}
\end{figure}

\begin{figure}[H]
  \centering
  \includegraphics[width=0.3\linewidth,clip]{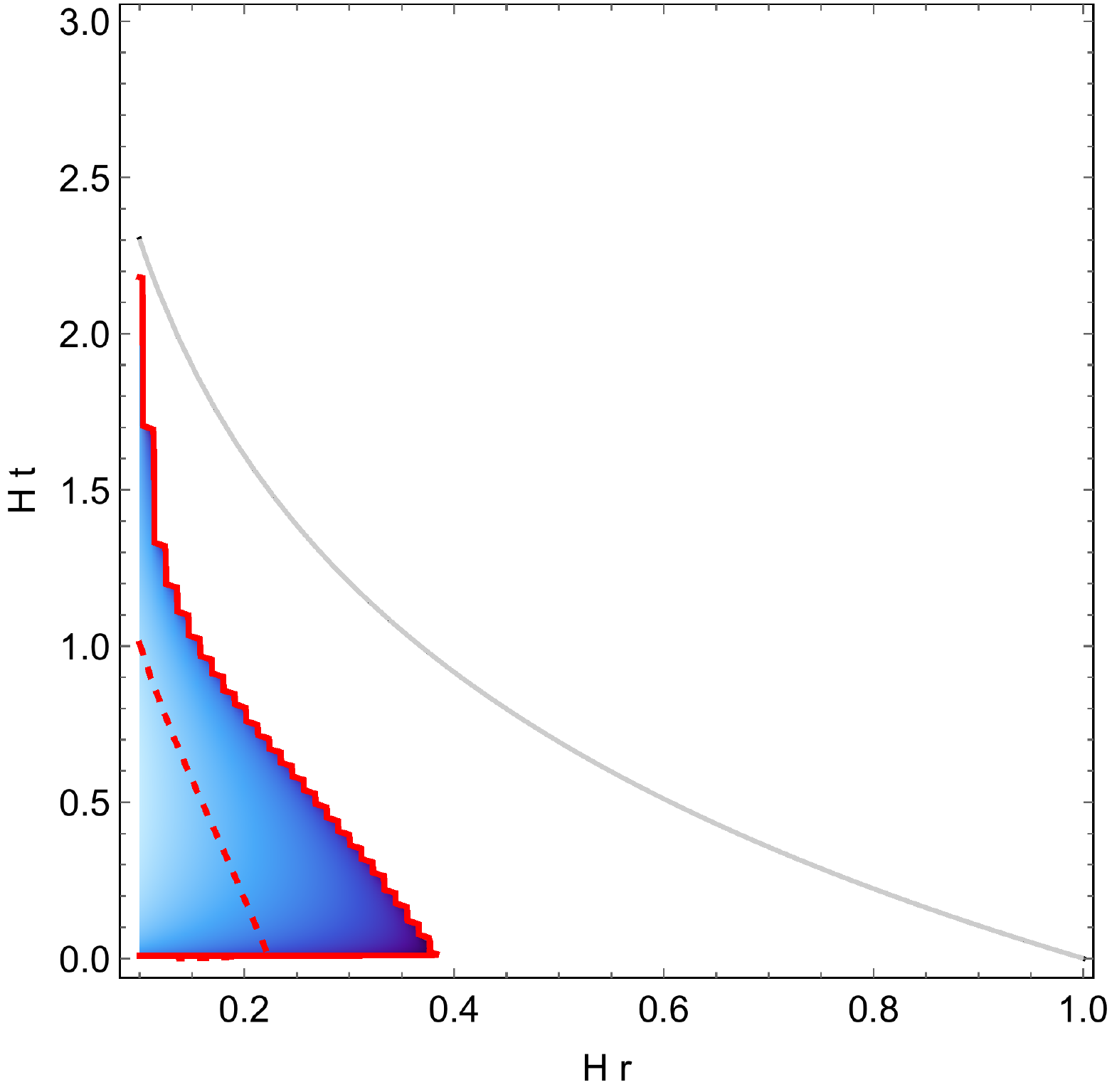}
  \includegraphics[width=0.3\linewidth,clip]{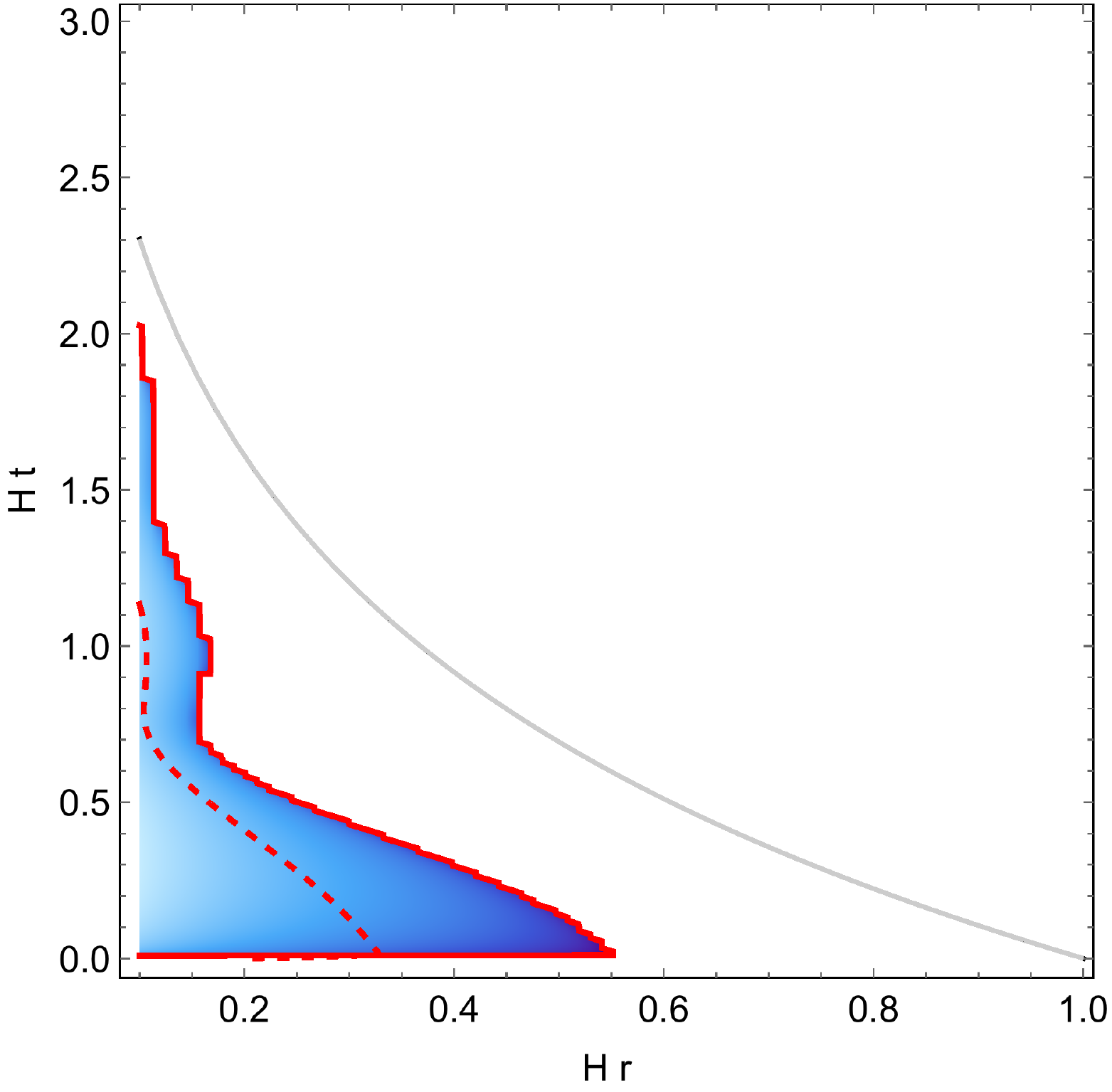}
  \includegraphics[width=0.3\linewidth,clip]{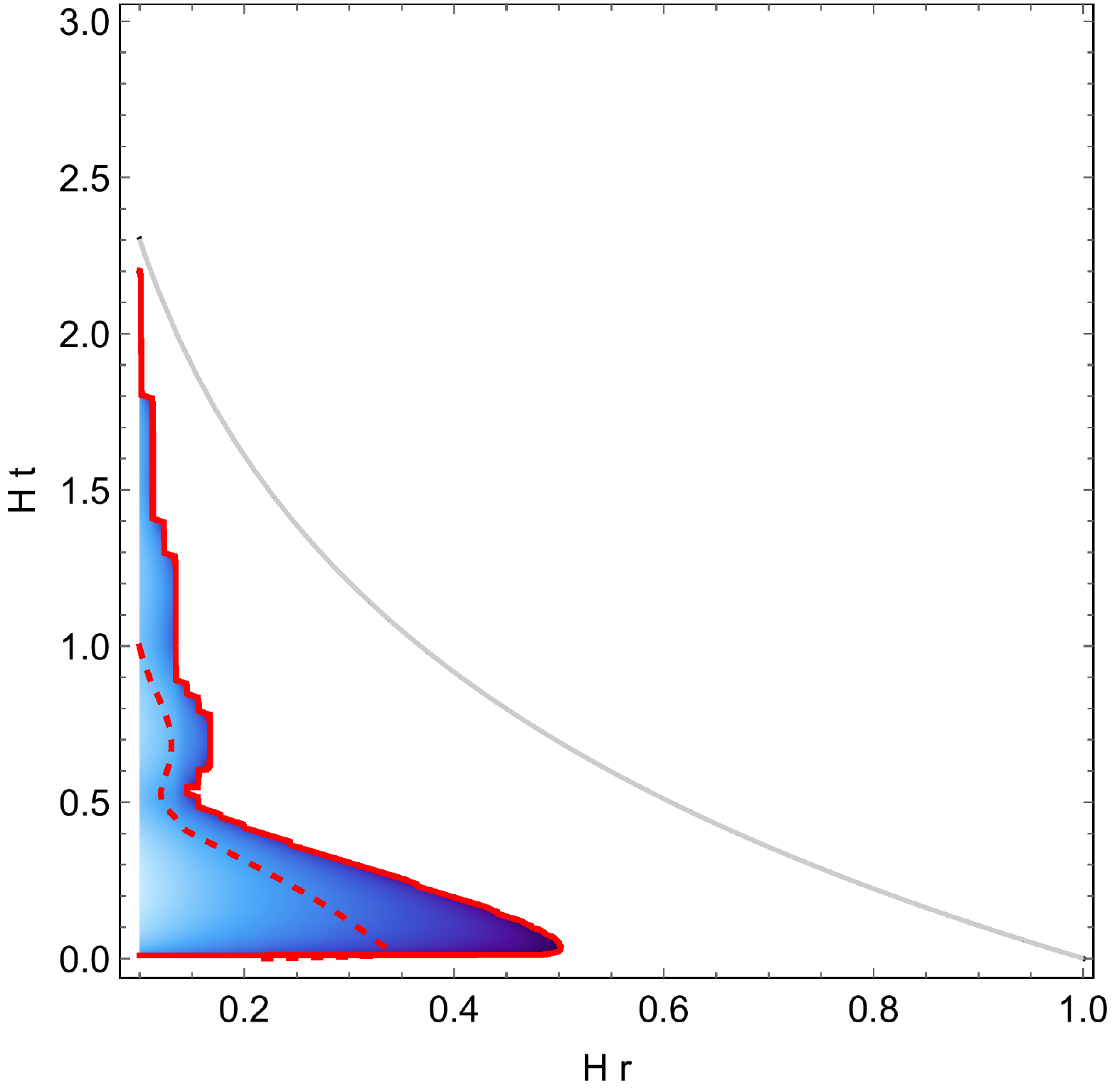}
  \caption{Evolution of negativity for between comoving detectors
    $\omega/H=2,4,6$ with $\lambda=0.1,\sigma H=0.4$. Detectors' world lines are
    $r=\text{const.}$ The gray solid line represents comoving size of
    the Hubble horizon and the red solid lines represent negativity
    zero contours.  Detectors are entangled for parameters in regions
    enclosed by the solid red lines. For parameters in  regions
    enclosed by the red dotted line, violation of Bell's inequality
    can be detected. }
\label{fig:neg-mini}
\end{figure}

\begin{figure}[H]
  \centering
    \includegraphics[width=0.4\linewidth,clip]{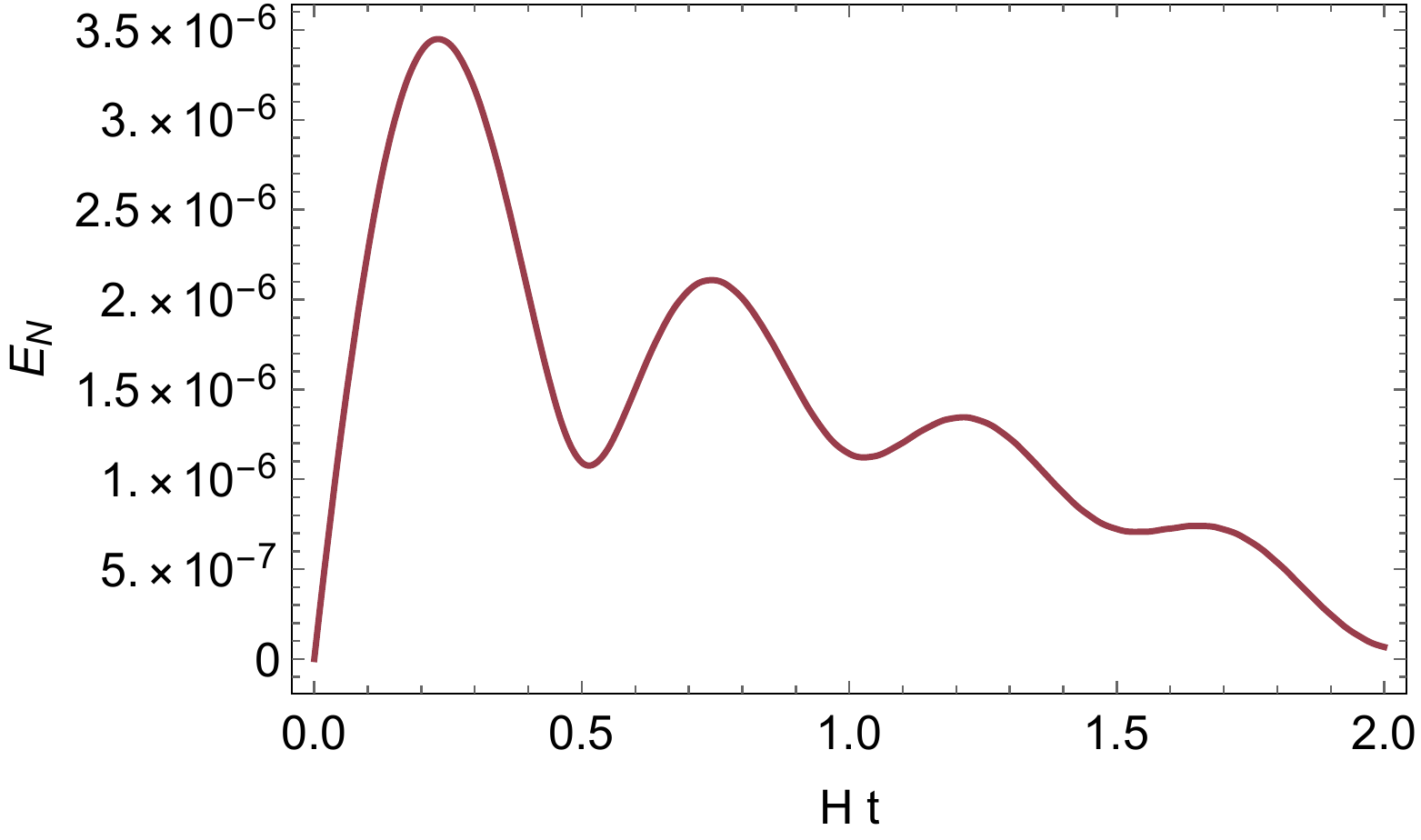}
  \includegraphics[width=0.4\linewidth,clip]{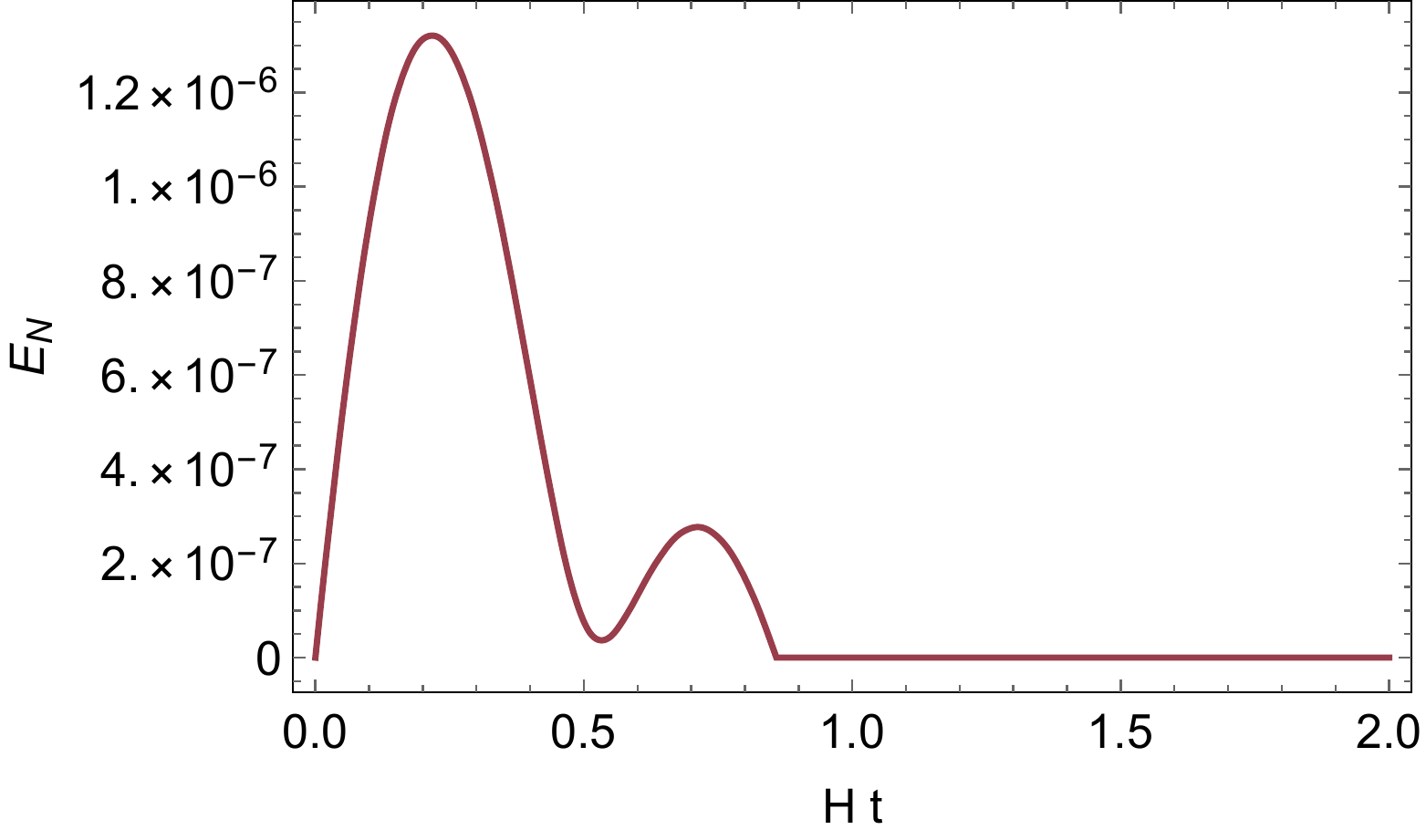}
  \caption{Evolution of negativity for $\lambda=0.1,\omega/H=6$ at $rH=0.1$ (left)
    and at $rH=0.2$ (right).}
\label{fig:mini-negt}
\end{figure}

Initial detection of entanglement determined by Eq.~\eqref{eq:neg0}
(Fig.~\ref{fig:mini-neg0}) . Compared to the conformal scalar case,
the maximum distance of the entanglement detection is reduced and
detection is possible only for $r<0.6H^{-1}$ (sub horizon scale).
Evolution of negativity is shown in Fig.~\ref{fig:neg-mini} and
Fig.~\ref{fig:mini-negt}. The behavior is almost same as that of the
conformal scalar field apart from initial spatial scale of
entanglement detection. As a result of evolution, for any values of
$r$, detected entanglement of the scalar field vanishes before
separation between two detectors exceeds the Hubble horizon
scale. This behavior is contrasted with the result of the conformal
scalar field, in which case the entanglement can extend beyond the
Hubble horizon for sufficiently small $r$. For the minimal scalar
field, particle creations  in de Sitter space  works as
a noise and destroys the quantum coherence over the Hubble scale. As
a result, the detectors can not catch the entanglement of the quantum
field for the scale larger than the Hubble horizon.
\noindent

\begin{figure}[H]
  \centering
  \includegraphics[width=0.3\linewidth,clip]{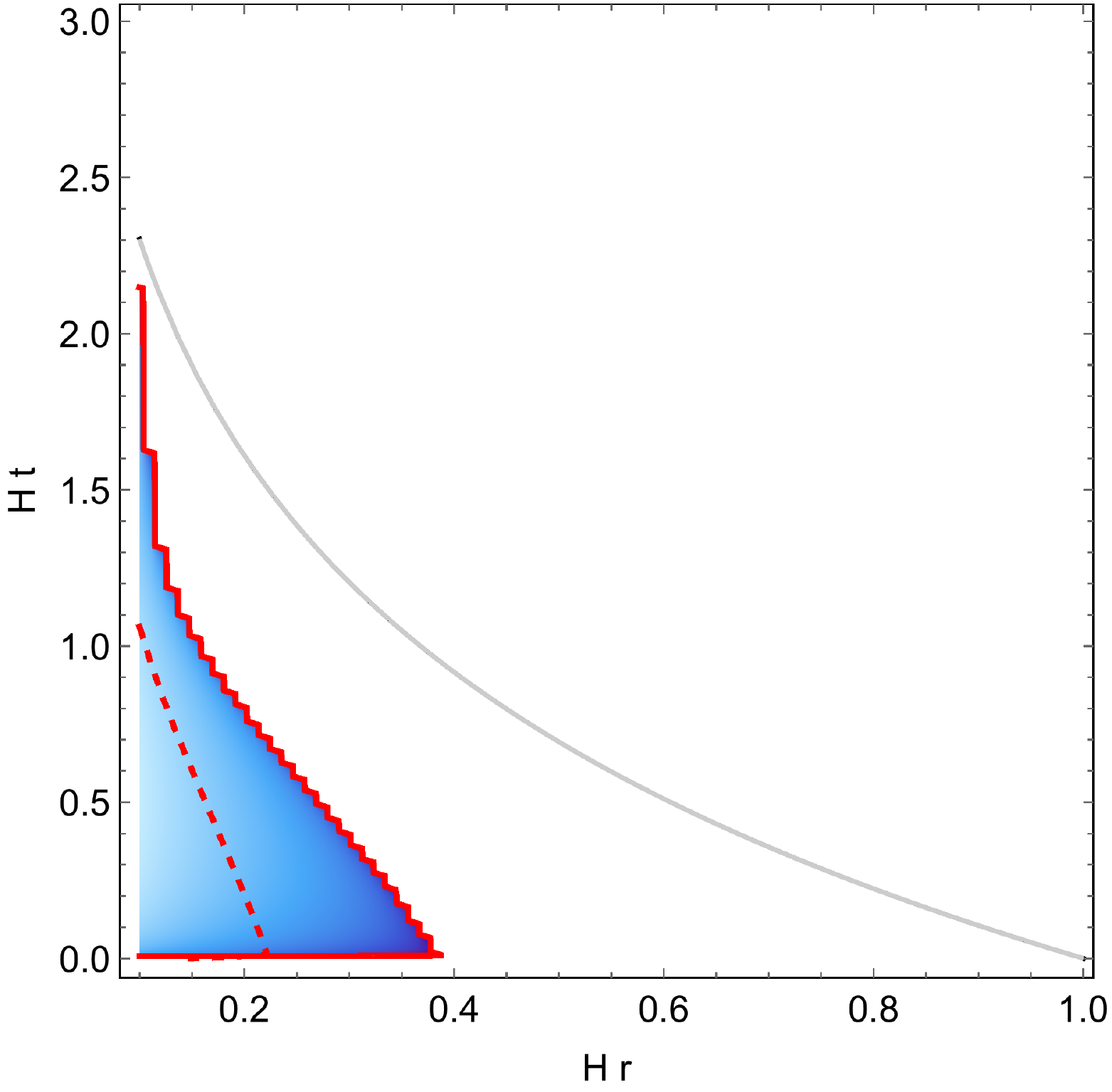}
  \includegraphics[width=0.3\linewidth,clip]{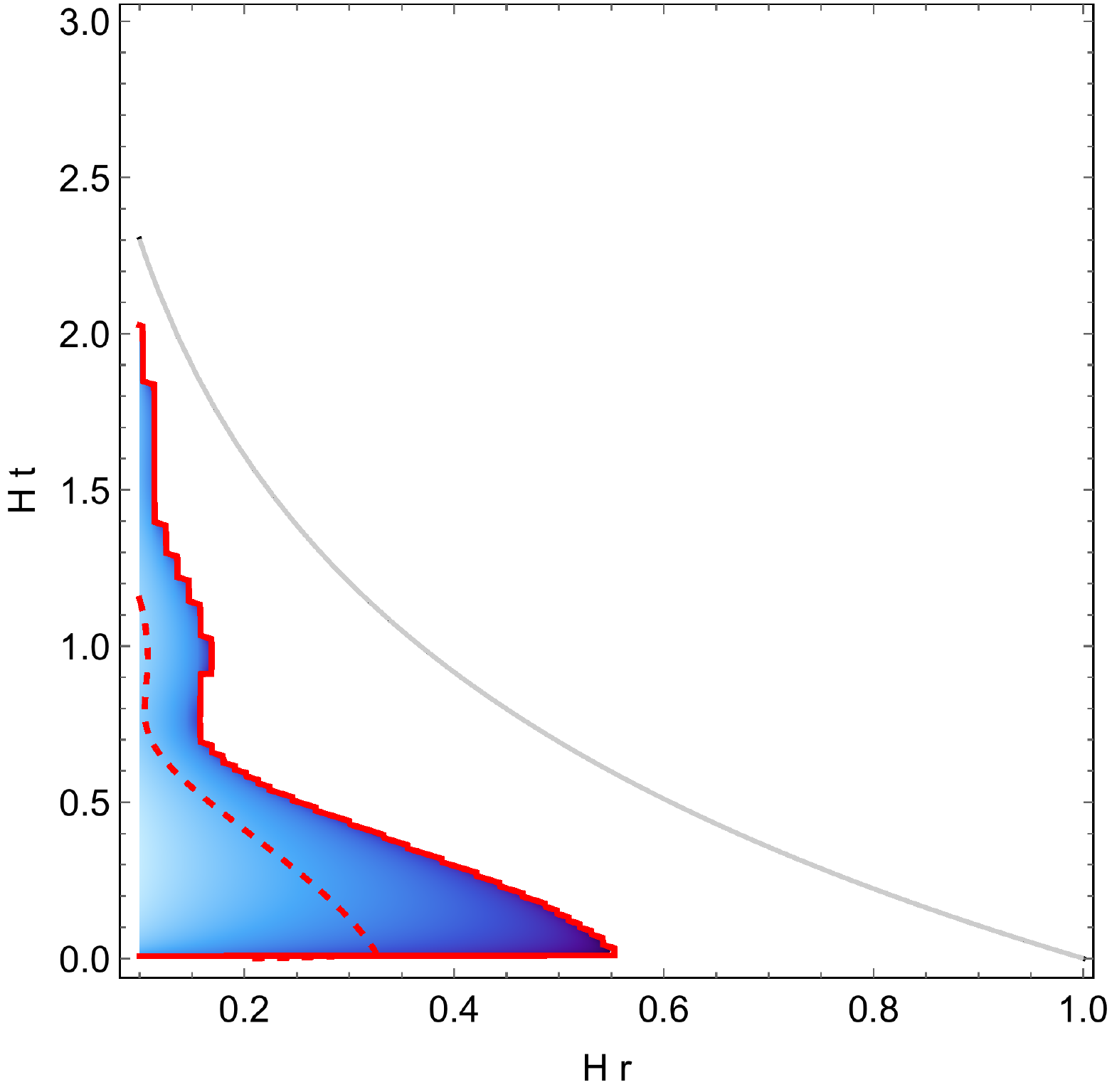}
  \includegraphics[width=0.3\linewidth,clip]{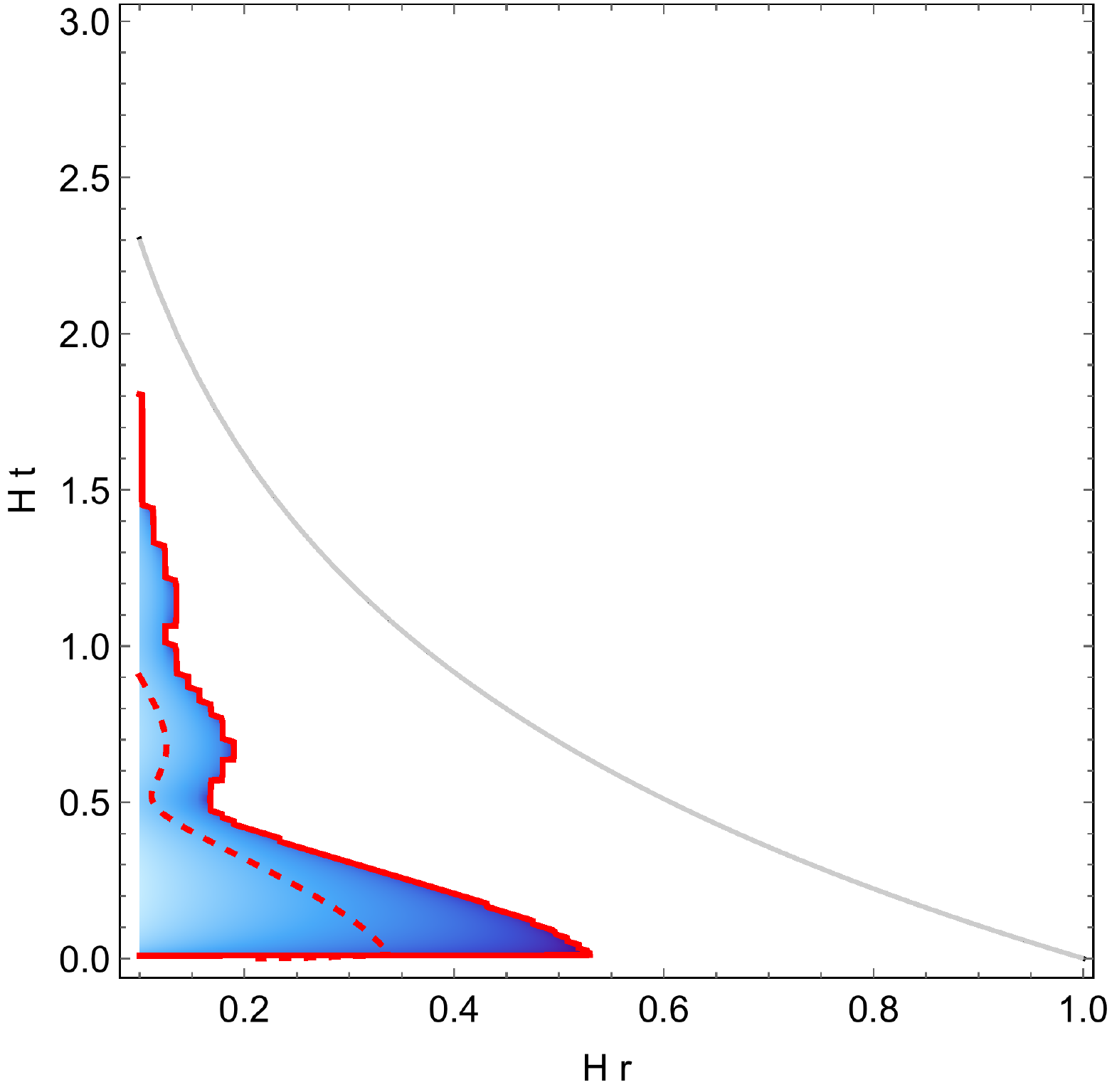}
  \caption{Evolution of negativity for between comoving detectors
    $\omega/H=2,4,6$ with $\lambda=0.5,\sigma H=0.4$. Detectors' world lines are
    $r=\text{const.}$ The gray solid line represents comoving size of
    the Hubble horizon and the red solid lines represent negativity
    zero contours.  Detectors are entangled for parameters in regions
    enclosed by the solid red lines. For parameters in  regions
    enclosed by the red dotted line, violation of Bell's inequality
    can be detected. }
\label{fig:neg-minib}
\end{figure}

\begin{figure}[H]
  \centering
    \includegraphics[width=0.4\linewidth,clip]{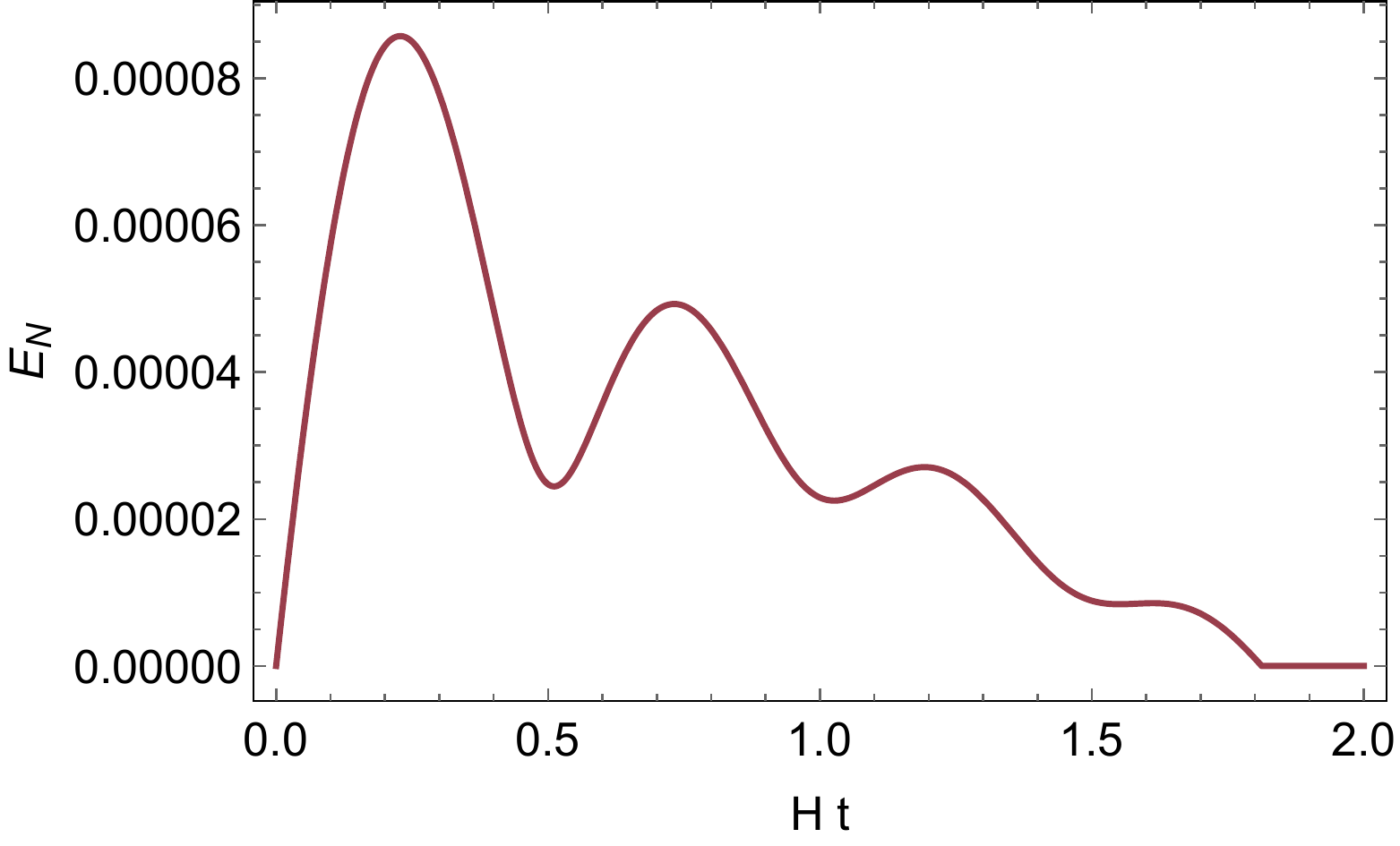}
  \includegraphics[width=0.4\linewidth,clip]{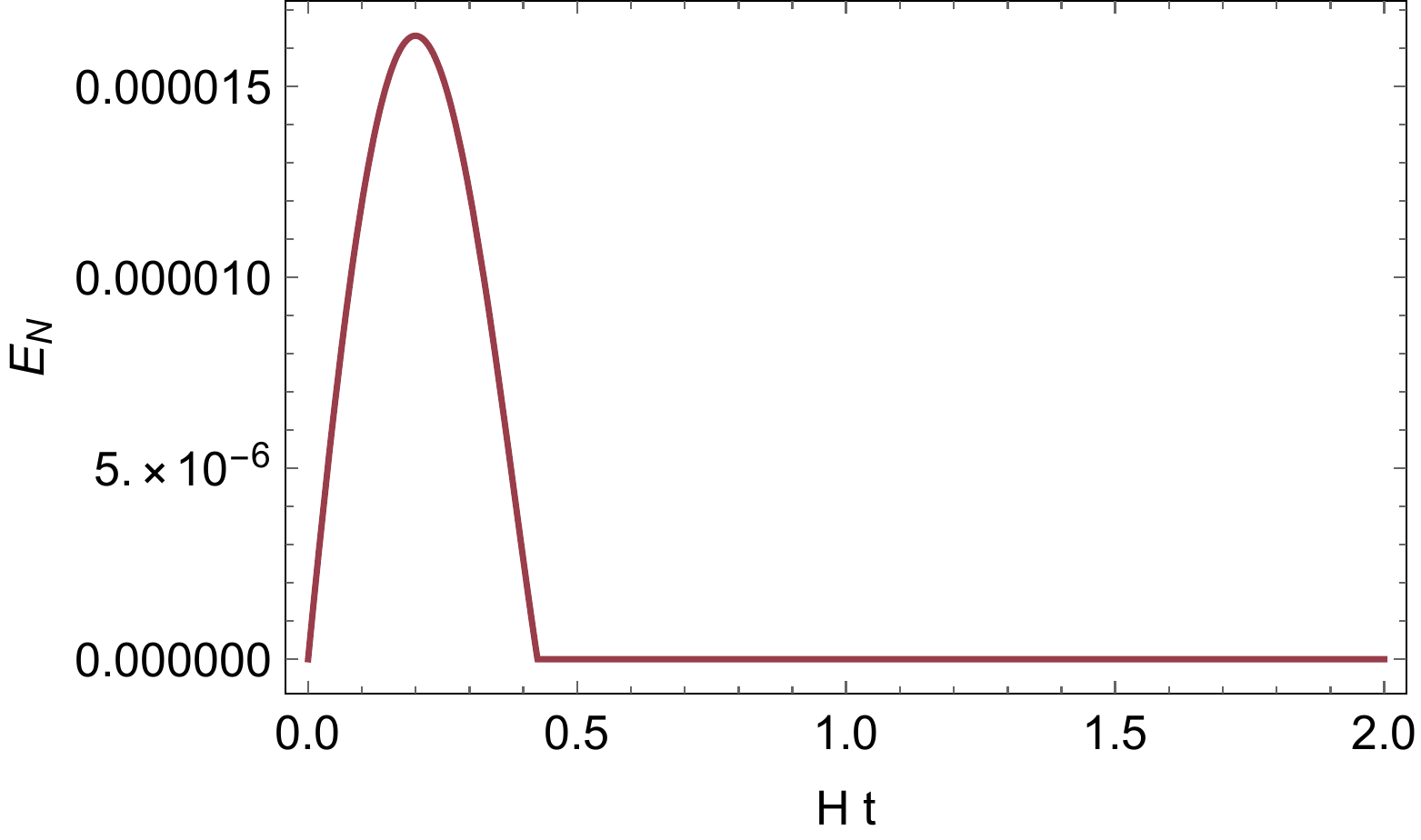}
  \caption{Evolution of negativity for $\lambda=0.5,\omega/H=6$ at $rH=0.1$ (left)
    and at $rH=0.2$ (right).}
\label{fig:mini-negtb}
\end{figure}

Fig.~\ref{fig:neg-minib} and Fig.~\ref{fig:mini-negtb} show evolution
of negativity for $\lambda=0.5$.
\noindent

\section{Summary}

We investigated evolution of entanglement between two detectors in de
Sitter space.  We adopt the master equation with DCG approximation to
obtain evolution of the negativity between two detectors interacting
with the quantum scalar field.  For the massless conformal scalar
field with the conformal vacuum, after evolution, two detectors can
catch the entanglement of the scalar field beyond the horizon scale if
the initial comoving separation is sufficiently small. At the same
time, violation of the Bell-CHSH inequality is detectable for scales
larger than the Hubble horizon. This behavior cannot be obtained
previous analysis based on naive
perturbation~\cite{Nambu2011,Nambu2013}.  On the other hand, for the
massless minimal scalar field with the Bunch-Davies vacuum, the
detected negativity decays within a Hubble time scale and two
detectors become separable before their physical distance exceeds the
Hubble horizon.  This behavior is consistent with the result obtained
by considering bipartite entanglement between two spatial regions
introduced by averaging~\cite{Nambu2008,Nambu2009}.  Thus the result
obtained in this paper supports appearance of classical nature of
fluctuations in the inflationary universe.

We should comment on relation to our recent lattice simulation of
negativity in de Sitter space~\cite{Matsumura2017}. We discretize the
minimal scalar field in de Sitter space and evaluated negativity
between two spatial regions in the 1-dim lattice. The analysis shows
that the entanglement on the super horizon scale always remains. This
result must be contrasted with the result obtained by the master
equation in this paper, which shows detectors cannot catch
entanglement beyond the Hubble horizon scale.  At this stage, we can
only say that detection of entanglement using a pair of two detectors
(bipartite entanglement in the present case) is not so effective and
there may be a possibility that the super horizon scale entanglement
can be detectable by considering effect of the multi-partite
entanglement. We will report on this subject in our next
publication~\cite{Kukita2017b}.

\begin{acknowledgments}
This work was supported in part by the JSPS KAKENHI Grant Number 16H01094.
\end{acknowledgments}


\end{document}